\documentclass[aip,amsmath,amssymb,groupedaddress]{revtex4-1}
\usepackage{graphicx} % arXiv に投稿するときは dvipdfmx オプションを外す
\usepackage[hang,small,bf]{caption}
\usepackage[subrefformat=parens]{subcaption}
\usepackage{braket}
\usepackage{hhline}
\everymath{\displaystyle}
\newcommand{\nunit}[2]{{#1}\;\mathrm{#2}}
\newcommand{\norm}[1]{\left\lVert{#1}\right\rVert}

\begin{document}
\title{Energy Spectrum Analysis on a Red Blood Cell Model}

\author{Tetsuya Yamamoto}
\email{tyamamoto-st@keio.jp}

\author{Hiroshi Watanabe}

\affiliation{
    Department of Applied Physics and Physico-Informatics, Faculty of Science and Technology, Keio University, Yokohama, Kanagawa 223-8522, Japan
}

\begin{abstract}
    It is important to understand the dynamics of red blood cells (RBCs) in blood flow. This requires the formulation of coarse-grained RBC models that reproduce the hydrodynamic properties of blood accurately. One of the models that successfully reproduce the rheology and morphology of blood has been proposed by Fedosov \textit{et al.} [D.~A.~Fedosov, B.~Caswell, and G.~E.~Karniadakis, \textit{Comput.~Methods Appl.~Mech.~Eng.} \textbf{199}, 1937--1948 (2010)]. The proposed RBC model contains several parameters whose values are determined either by various experiments or physical requirements. In this study, we developed a new method of determining the parameter values precisely from the fluctuations of the RBC membrane. Specifically, we studied the relationship between the spectra of the fluctuations and model parameters. Characteristic peaks were observed in the spectra, whose peak frequencies were dependent on the parameter values. In addition, we investigated the spectra of the radius of gyration. We identified the peaks originating from the spring potential and the volume-conserving potential appearing in the spectra. These results lead to the precise experimental determination of the parameters used in the RBC model.
\end{abstract}

\maketitle

\section{Introduction} \label{sec:introduction}

Blood has important functions in the human body such as carrying oxygen and nutrients. In recent years, blood is increasingly regarded as a promising medium to transport drugs using micro- and nano-carriers\cite{shi2017}. The main component of blood is the red blood cells (RBCs), whose properties change in the presence of diseases such as malaria and sickle cell disease\cite{liu1991,park2008}. Therefore, understanding the properties of RBCs in blood flow is crucial in the diagnosis and treatment of diseases, as well as the design of efficient drug carriers. Research in this field had been limited to experimental and theoretical studies. However, numerical methods capable of simulating blood flow were successively developed in the 1990s and thereafter\cite{hoogerbrugge1992,espanol1995,mcnamara1988,malevanets1999,takeda1994,fedosov2014,ye2016}. This opened up new ways to study the rheology, morphology, and dynamics of RBCs.

Among the most adopted numerical methods of simulating RBCs are dissipative particle dynamics (DPD)\cite{hoogerbrugge1992,espanol1995}, the lattice Boltzmann method (LBM)\cite{mcnamara1988}, multiparticle collision dynamics (MPC)\cite{malevanets1999}, and smoothed particle hydrodynamics (SPH)\cite{takeda1994} simulations. We refer the readers to some reviews for further details on these methods\cite{fedosov2014,ye2016}. In DPD simulations of RBCs, for the degrees of freedom of all cell structures---the cell membrane, cytoskeleton, cytoplasm, and blood plasma---the structures are treated as Lagrangian particles. This approach enables us to flexibly and straightforwardly simulate complex fluids, keeping the conservation laws of hydrodynamics\cite{ye2016}.

In DPD models of RBCs, several parameters are determined to link the macroscopic properties of RBCs with the microscopic properties of DPD particles. Macroscopic quantities of the RBC membrane include the shear modulus and bending rigidity, which have been measured experimentally using optical tweezers and by atomic force microscopy and micropipette aspiration\cite{lenormand2001,scheffer2001,mohandas1994,matthews2022}. However, in the case of the shear modulus for instance, its values have been determined to be $\nunit{4\,\text{--}\, 9}{\mu N/m}$ using optical tweezers, whereas micropipette aspiration experiments have yielded values of $\nunit{5\,\text{--}\, 12}{\mu N/m}$\cite{fedosov2010}. Because of the limited accuracy of experimental measurements, the model parameters cannot be determined accurately. In addition, the discretization of the RBC membrane further causes the uncertainty of the model parameters.

To address the above issues, we propose a new method of determining the model parameters precisely from the fluctuations of the RBC membrane. First observed in the 19th century, the fluctuations of the RBC membrane have been extensively studied especially with regard to their origins\cite{gnesotto2018}. Thermal fluctuations have been studied in terms of the membrane displacement and its Fourier spectrum by G\"{o}gler \textit{et al.}\cite{gogler2007} On the other hand, Turlier \textit{et al.}\cite{turlier2016} found that non-equilibrium fluctuations due to cell metabolism violate the fluctuation--dissipation relation. Interferometric optical tweezer techniques employed in these studies enable the measurement of RBC membrane fluctuations with sub-nanometer precision in the frequency range of $\nunit{10^{-1}\,\text{--}\, 10^5}{Hz}$. Once measured, the membrane fluctuations can be analyzed in the frequency domain through Fourier transforms. In the case of a DPD model of RBCs, the Fourier spectra of the fluctuations are dependent on the model parameters governing the membrane properties. In the present study, we adopted the DPD model proposed by Fedosov \textit{et al.}\cite{fedosov2010} to study its fluctuation-induced spectra in detail. We measured the Fourier spectra of the fluctuations of potential energies and identified which peaks in the spectra originated from which parameters. We also measured the spectra of membrane displacement, which is measurable by experiments. These steps provide a numerical basis with which experimental results can be compared to determine the DPD model parameters.

The rest of the paper is organized as follows. In Sec.~\ref{sec:method} , we describe the model of the RBC membrane. The results are described in Sec.~\ref{sec:results}. Sec.~\ref{sec:summary} is devoted to the summary and discussion.

\section{Method} \label{sec:method}

\subsection{Modeling of a single RBC} \label{subsec:rbc_model}

\subsubsection{Shaping of the RBC membrane} \label{subsubsec:rbc_membrane}

The shape of the RBC membrane was determined as follows. First, each face of a regular icosahedron was split into smaller regular triangles, and the resulting vertices were moved radially to the surface of a sphere. Each vertex was then stereographically projected onto a plane. The vertices were subjected to Delaunay triangulation to obtain a list of points (particles) comprising each regular triangle. Finally, the sphere was formed into the shape of an RBC membrane according to the following equation\cite{fung1981}:
\begin{equation}
    \label{eq:rbcsurf}
    z={\pm}D_0\sqrt{1-\frac{4\left(x^2+y^2\right)}{D_0^2}}\left[a_0+a_1\frac{x^2+y^2}{D_0^2}+a_2\frac{\left(x^2+y^2\right)^2}{D_0^4}\right].
\end{equation}
Here, $D_0=\nunit{7.82}{\mu m}$ is the diameter of the RBC, and the constants are $a_0=0.0518$, $a_1=2.0026$, and $a_2=-4.491$. We chose the number of particles $N=492$ from the fact that the RBC membrane is widely discretized using $N=500$ in various simulations\cite{fedosov2010,fedosov2011,peng2013}. As noted in Appendix~\ref{sec:coarse_graining}, we confirm that the degree of coarse graining, i.e., the number of membrane particles, has little effect on the observed peak frequencies.

\begin{figure}[htbp]
    \begin{subfigure}[b]{.49\textwidth}
        \vspace{0.5cm}
        \centering
        \includegraphics[trim={0cm 0cm 0cm 0cm},clip,height=.7\linewidth]{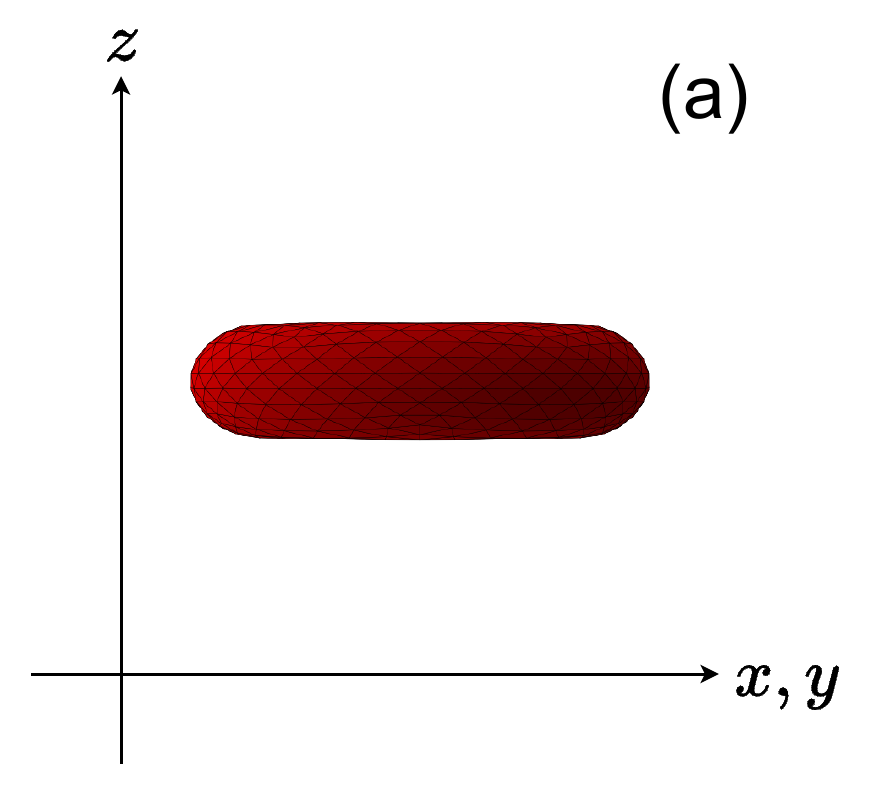}
        \label{subfig:rbc_side}
    \end{subfigure}
    \begin{subfigure}[b]{.49\textwidth}
        \centering
        \includegraphics[trim={0cm 0cm 0cm 0cm},clip,height=.7\linewidth]{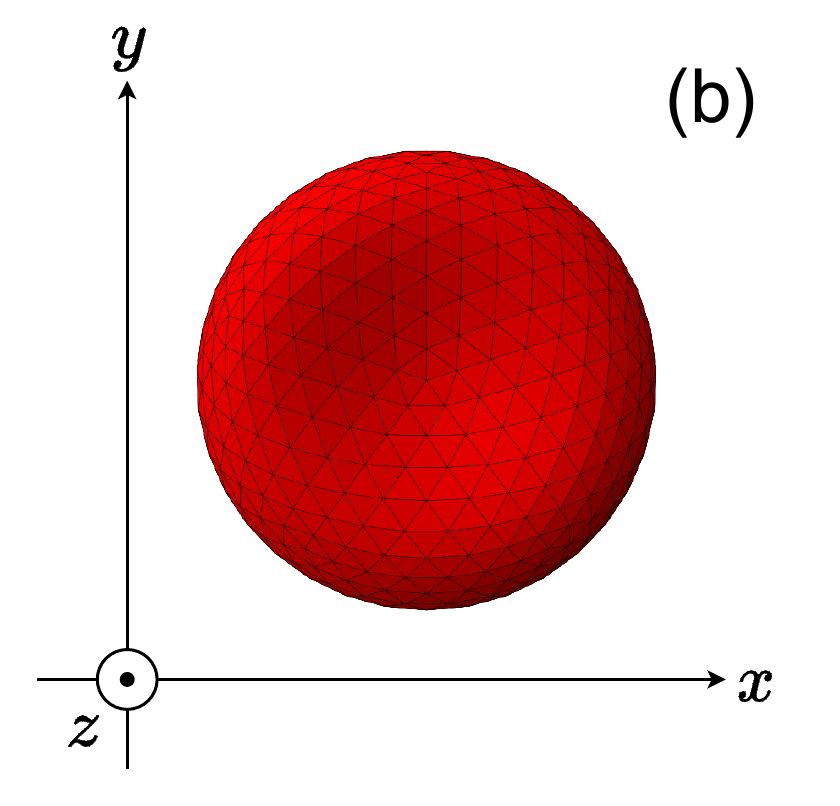}
        \label{subfig:rbc_above}
    \end{subfigure}
    \caption{(Color online) Orientation of the RBC in Cartesian coordinates. (a) Side and (b) top views of the RBC.}
    \label{fig:rbc}
\end{figure}

Figure~\ref{fig:rbc} shows the orientation of the RBC in Cartesian coordinates. We chose the $z$-axis as the axis of rotational symmetry of the RBC and the $xy$-plane perpendicular to it.

\subsubsection{RBC membrane potentials} \label{subsubsec:rbc_potentials}
In this section, we consider the potentials governing the membrane particles. The total potential energy $V_\mathrm{RBC}$ is written as
\begin{equation}
    \label{eq:Vrbc}
    V_\mathrm{RBC} = V_\mathrm{spring} + V_\mathrm{area} + V_\mathrm{volume} + V_\mathrm{bending}.
\end{equation}
In the following, we will outline the individual potentials. See the paper by Fedosov \textit{et al.} for more details\cite{fedosov2010}. The values of the parameters involving the potentials are listed in Table \ref{tab:parameters} at the end of this section.

First, $V_\mathrm{spring}$ acts on the two particles making up each side of a triangular lattice. This potential corresponds to spring forces that respond to external stress and is given by
\begin{equation}
    \label{eq:Vspring}
    V_\mathrm{spring}=\underbrace{-\sum_{j=1}^{N_\mathrm{s}}\frac{k_\mathrm{s}}{2}(l_j^\mathrm{m})^2\log{(1-x_j^2)}}_{V_\mathrm{FENE}}+\underbrace{\sum_{j=1}^{N_\mathrm{s}}\dfrac{k_\mathrm{p}^j}{l_j}}_{V_\mathrm{POW}},
\end{equation}
where the first term is the finitely extensible nonlinear elastic (FENE) potential $V_\mathrm{FENE}$\cite{kremer1990}. This potential yields attractive forces, whereas the second-term $V_\mathrm{POW}$ is a repulsive potential. $N_\mathrm{s}$ is the number of springs, $l_j$ is the length of the $j$th spring, $l_j^\mathrm{m}$ is its maximum length, and $x_j=l_j/l_j^\mathrm{m}$. $k_\mathrm{s}$ is a constant set to the same value for all springs. $k_\mathrm{p}^j$ is set individually for each spring so that the spring forces cancel out at the equilibrium length $l_j^0$. In this study, $l_j^0$ was set equal to $l_j$ at the RBC's initial state as shown in Fig.~\ref{fig:rbc}. Then, $l_j^\mathrm{m}$ was determined by fixing the ratio $x_0=l_j^0/l_j^\mathrm{m}$ for all springs. The values of $k_\mathrm{s}$ and $k_\mathrm{p}^j$ are determined from the membrane shear modulus measured by experiments\cite{fedosov2010,li2005,dao2006}.

The RBC membrane is nearly incompressible; it maintains constant surface area and volume regardless of its shape\cite{gompper2008}. This is reflected in the model through the area-conserving potential $V_\mathrm{area}$ and the volume-conserving potential $V_\mathrm{volume}$. The potentials are written as
\begin{align}
    V_\mathrm{area}   & =\frac{k_\mathrm{a}\left(A_t^\mathrm{tot}-A_0^\mathrm{tot}\right)^2}{2A_0^\mathrm{tot}}, \label{eq:Varea}   \\
    V_\mathrm{volume} & =\frac{k_\mathrm{v}\left(V_t^\mathrm{tot}-V_0^\mathrm{tot}\right)^2}{2V_0^\mathrm{tot}}, \label{eq:Vvolume}
\end{align}
where $k_\mathrm{a}$ and $k_\mathrm{v}$ are constants, $A_0^\mathrm{tot}$ is the initial area, and $V_0^\mathrm{tot}$ is the initial volume. $A_t^\mathrm{tot}$ and $V_t^\mathrm{tot}$ are the area and volume at time $t$, which are constrained to $A_0^\mathrm{tot}$ and $V_0^\mathrm{tot}$, respectively. The coefficients $k_\mathrm{a}$ and $k_\mathrm{v}$ are respectively set to $\nunit{315}{\mu N/m}$ and $\nunit{1.23}{kPa}$, which are large enough to provide a nearly incompressible membrane.

The characteristic biconcave shape of the RBC is maintained by the bending energy of the membrane\cite{gompper2008}. It is introduced in the model as $V_\mathrm{bending}$ given by
\begin{equation}
    \label{eq:Vbending}
    V_\mathrm{bending}=\sum_{j=1}^{N_\mathrm{s}} k_\mathrm{b}\left[1-\cos{\left(\theta_j-\theta_0\right)}\right],
\end{equation}
where $k_\mathrm{b}$ is a constant and $\theta_j$ is the dihedral angle between the two triangular lattices sharing the $j$th edge. $\theta_0$ is the spontaneous angle, which is often set to $\theta_0=0$ in the literature as is the case in this study\cite{turlier2016,gompper1997,peng2013}. The coefficient $k_\mathrm{b}$ is derived from experimental measurements of the bending rigidity of the membrane\cite{fedosov2010,li2005,dao2006}.

From the potentials defined above, the conservative force $\boldsymbol{F}_i^\mathrm{C}$ acting on the $i$th particle is given by
\begin{align}
    \boldsymbol{F}_i^\mathrm{C}
     & =-\boldsymbol{\nabla}_i V_\mathrm{RBC} \nonumber                                                            \\
     & =-\boldsymbol{\nabla}_i\left(V_\mathrm{spring}+V_\mathrm{area}+V_\mathrm{volume}+V_\mathrm{bending}\right).
\end{align}
The equations of each nodal force are described in Appendix \ref{sec:forces}.

\subsection{Dissipative particle dynamics} \label{subsec:DPD}

In this study, we adopted dissipative particle dynamics (DPD) to regulate the temperature of the RBC model. DPD is a stochastic thermostat similar to the Langevin thermostat, although the former conserves the total translational and angular momenta. The equations of motion for the $i$th DPD particle are written as\cite{espanol1995}
\begin{equation}
    \label{eq:dpd}
    \begin{aligned}
         & m\dot{\boldsymbol{v}}_i=\sum_{j\neq i} \boldsymbol{F}_{ij}^\text{C}+\sum_{j\neq i} \boldsymbol{F}_{ij}^\text{D}+\sum_{j\neq i} \boldsymbol{F}_{ij}^\text{R}, \\
         & \begin{cases}
                & \boldsymbol{F}_{ij}^\text{D}=-\gamma\omega^\text{D}(r_{ij})(\boldsymbol{v}_{ij}\cdot \boldsymbol{e}_{ij})\boldsymbol{e}_{ij}, \\
                & \boldsymbol{F}_{ij}^\text{R}=\sigma\omega^\text{R}(r_{ij})\xi_{ij}\boldsymbol{e}_{ij},
           \end{cases}
    \end{aligned}
\end{equation}
where $\boldsymbol{F}_{ij}^\text{C},\ \boldsymbol{F}_{ij}^\text{D}$, and $\boldsymbol{F}_{ij}^\text{R}$ are the conservative, dissipative, and random forces, respectively. Furthermore, $\boldsymbol{r}_{ij}=\boldsymbol{r}_i-\boldsymbol{r}_j,\ r_{ij}=\norm{\boldsymbol{r}_{ij}},\ \boldsymbol{e}_{ij}=\boldsymbol{r}_{ij}/r_{ij}$, and $\boldsymbol{v}_{ij}=\boldsymbol{v}_i-\boldsymbol{v}_j$. A white-noise term following the standard normal distribution is denoted by $\xi_{ij}$. The relations $\xi_{ij}=\xi_{ji}$, $\boldsymbol{e}_{ij}=-\boldsymbol{e}_{ji}$, and $\boldsymbol{v}_{ij}=-\boldsymbol{v}_{ji}$ guarantee the conservation of momenta. Weight functions are denoted by $\omega^\text{D}(r_{ij})$ and $\omega^\text{R}(r_{ij})$, whereas $\gamma$ and $\sigma$ are constant coefficients. The weight functions and coefficients independently satisfy Einstein's relation\cite{espanol1995} through
\begin{equation}
    \label{eq:dpdeinstein}
    \begin{aligned}
        \left[\omega^\text{R}(r_{ij})\right]^2 & =\omega^\text{D}(r_{ij}), \\
        \sigma^2                               & =2\gamma k_\text{B}T.
    \end{aligned}
\end{equation}

\subsection{Simulation details} \label{subsec:sim_details}

\subsubsection{Measurement} \label{subsubsec:measurement}

The RBC membrane was first equilibrated in the NVT ensemble, i.e., the isothermal condition, by regulating the membrane temperature using a DPD thermostat. After the membrane reached equilibrium, the thermostat was turned off, and simulations were performed in the NVE ensemble, i.e., the isoenergetic condition. For simulation results in the NVT ensemble, see Appendix~\ref{sec:dpd_on_spectra}.

The membrane was first thermalized for $\nunit{1.1}{ms}$ ($40000$ steps) with a DPD thermostat. Then, the thermostat was turned off and the simulation was performed in the NVE ensemble for another $\nunit{1.1}{ms}$, where the physical quantities were computed. The particle averages of the potentials $V_\mathrm{FENE},\ V_\mathrm{POW},\ V_\mathrm{area},\ V_\mathrm{volume}$, and $V_\mathrm{bending}$ were separately calculated at every step, to which the temporal Fourier transform $(t\to f=\omega/(2\pi))$ was applied. The Fourier spectra were computed for two different values of the parameters $k_\mathrm{s}$ of $V_\mathrm{FENE}$, $k_\mathrm{p}$ of $V_\mathrm{POW}$, $k_\mathrm{a}$ of $V_\mathrm{area}$, $k_\mathrm{v}$ of $V_\mathrm{volume}$, and $k_\mathrm{b}$ of $V_\mathrm{bending}$, that is, the original value and a value set $20\%$ smaller. When a peak shift was observed between the two different values of a parameter, the peak was identified as originating from the corresponding membrane potential. We performed 7000 independent runs for each parameter with different random seeds of the DPD thermostat.

Experiments cannot directly measure the potential energy of the membrane, unlike numerical calculations. Therefore, we also measured the radius of gyration $R_\alpha\ (\alpha=x,y,z)$ as an experimentally measurable quantity. $R_\alpha$ represents the spatial spread of membrane particles along the $\alpha$-axis and is defined as
\begin{equation}
    \label{eq:gyr_radius}
    R_\alpha=\sqrt{\sum_{i=1}^{N}\frac{\alpha_i^2}{N}},
\end{equation}
where $N$ is the number of particles. The Cartesian coordinates of the $i$th particle $(i=1,2,\cdots,N)$ relative to the center of mass of the RBC are written as $(x_i,y_i,z_i)$. Optical tweezer experiments can currently measure the RBC membrane displacement with sub-nanometer precision\cite{gogler2007}. This corresponds to the measurement of the radius of gyration in our simulations.

\subsubsection{Membrane mass} \label{subsubsec:membrane_mass}

The mass of the RBC membrane $M_\mathrm{memb}$ must be explicitly treated in this study as the frequency $f$ of the Fourier spectra is scaled as $f\propto\sqrt{k/M_\mathrm{memb}}$, where $k=k_\mathrm{s},\ k_\mathrm{a},\ k_\mathrm{v}$, or $k_\mathrm{b}$. The mean corpuscular hemoglobin---the total mass of the protein hemoglobin in a single RBC---has been measured to be $\nunit{28\,\text{--}\, 29}{pg}$\cite{kaza2021,moon2012}. Considering that hemoglobin composes $95\,\text{--}\, 98\%$ of the RBC mass without water, we treat the remaining mass ($2\,\text{--}\, 5\%$) to be that of the RBC membrane. This yields a membrane mass of $\nunit{0.6\,\text{--}\, 1.5}{pg}$. Here, we take it to be $\nunit{1.0}{pg}$, which is distributed evenly among the membrane particles.

\subsection{Model parameters} \label{sec:parameters}

Model parameters and their values are listed below in Table 1 in terms of the model and SI units.
\begin{table}[htbp]
    \caption{List of model parameters.}
    \label{tab:parameters}
    \centering
    \begin{tabular}{|c c c c|}
        \hline
        parameter                           & symbol          & \quad value (model units) & \quad value (SI units)
        \\ \hhline{|====|}
        number of steps                     & ---             & $80000$                   & $80000$
        \\ \hline
        time-step size                      & ---             & $0.005$                   & $\nunit{28}{ns}$
        \\ \hline
        cutoff radius                       & ---             & $12.2$                    & $\nunit{3.13}{\mu m}$
        \\ \hline
        damping coefficient                 & $\gamma$        & $0.05$                    & $\nunit{18}{pg/s}$
        \\ \hline
        temperature                         & $k_\mathrm{B}T$ & $1$                       & $\nunit{4.14\times 10^{-21}}{J}$
        \\ \hline
        particle mass                       & $m$             & $1$                       & $\nunit{2}{fg}$
        \\ \hline
        number of particles                 & $N$             & $492$                     & $492$
        \\ \hline
        RBC diameter                        & $D_0$           & $30.5$                    & $\nunit{7.82}{\mu m}$
        \\ \hline
        ---                                 & $a_0$           & $0.0518$                  & $0.0518$
        \\ \hline
        ---                                 & $a_1$           & $2.0026$                  & $2.0026$
        \\ \hline
        ---                                 & $a_2$           & $-4.491$                  & $-4.491$
        \\ \hline
        coefficient of $V_\mathrm{FENE}$    & $k_\mathrm{s}$  & $48.6$                    & $\nunit{3.06}{\mu N/m}$
        \\ \hline
        coefficient of $V_\mathrm{area}$    & $k_\mathrm{a}$  & $5000$                    & $\nunit{315}{\mu N/m}$
        \\ \hline
        coefficient of $V_\mathrm{volume}$  & $k_\mathrm{v}$  & $5000$                    & $\nunit{1.23}{kPa}$
        \\ \hline
        coefficient of $V_\mathrm{bending}$ & $k_\mathrm{b}$  & $66.9$                    & $\nunit{2.77\times 10^{-19}}{J}$
        \\ \hline
        spring eq.~length vs max.~length    & $x_0$           & $0.488$                   & $0.488$
        \\ \hline
    \end{tabular}
\end{table}

\section{Results} \label{sec:results}

\subsection{Time evolution of energies} \label{subsec:time_evo}

\begin{figure}[htbp]
    \begin{subfigure}[b]{.49\textwidth}
        \centering
        \includegraphics[width=\linewidth]{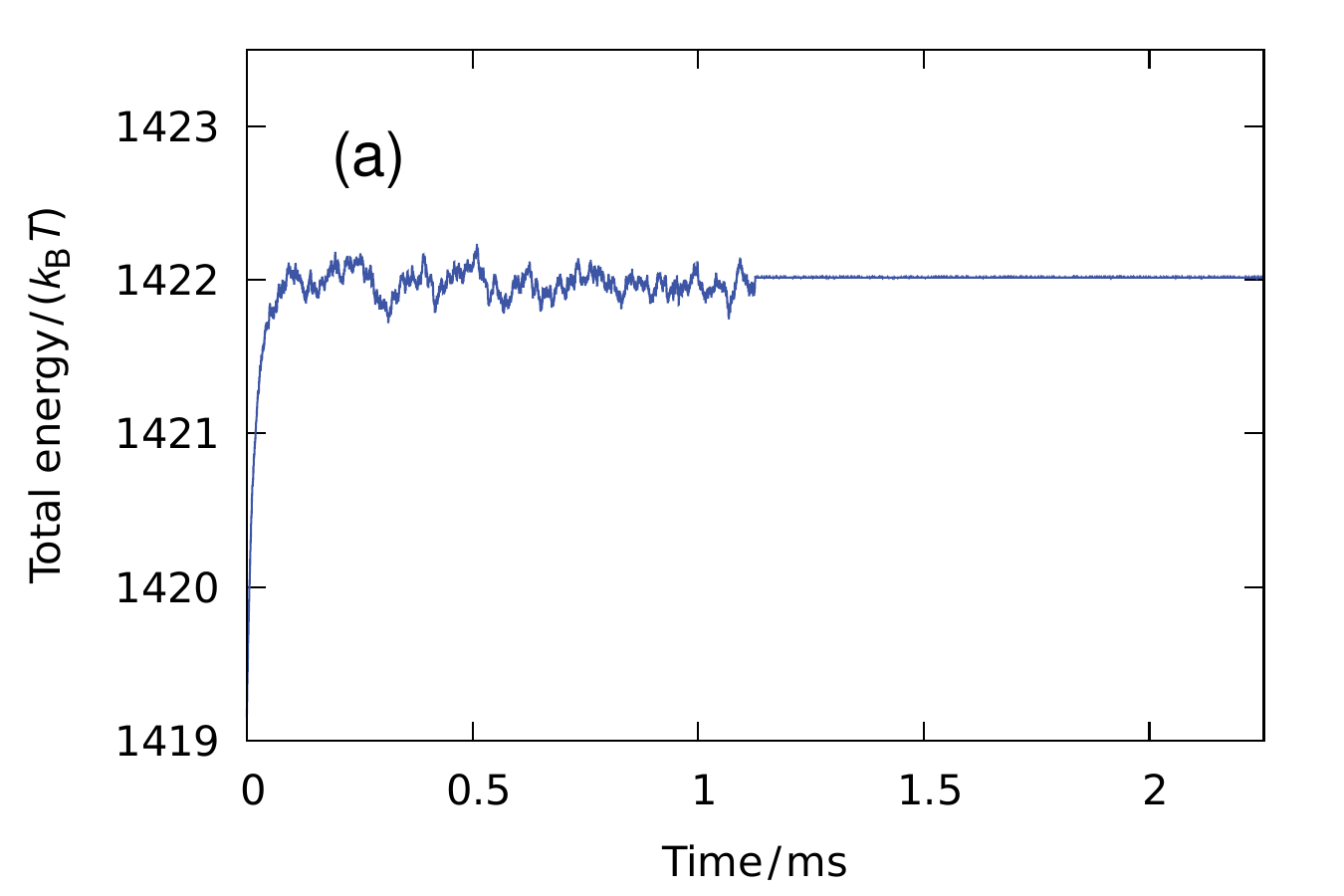}
        \label{subfig:totenergy}
    \end{subfigure}
    \hfill
    \begin{subfigure}[b]{.49\textwidth}
        \centering
        \includegraphics[width=\linewidth]{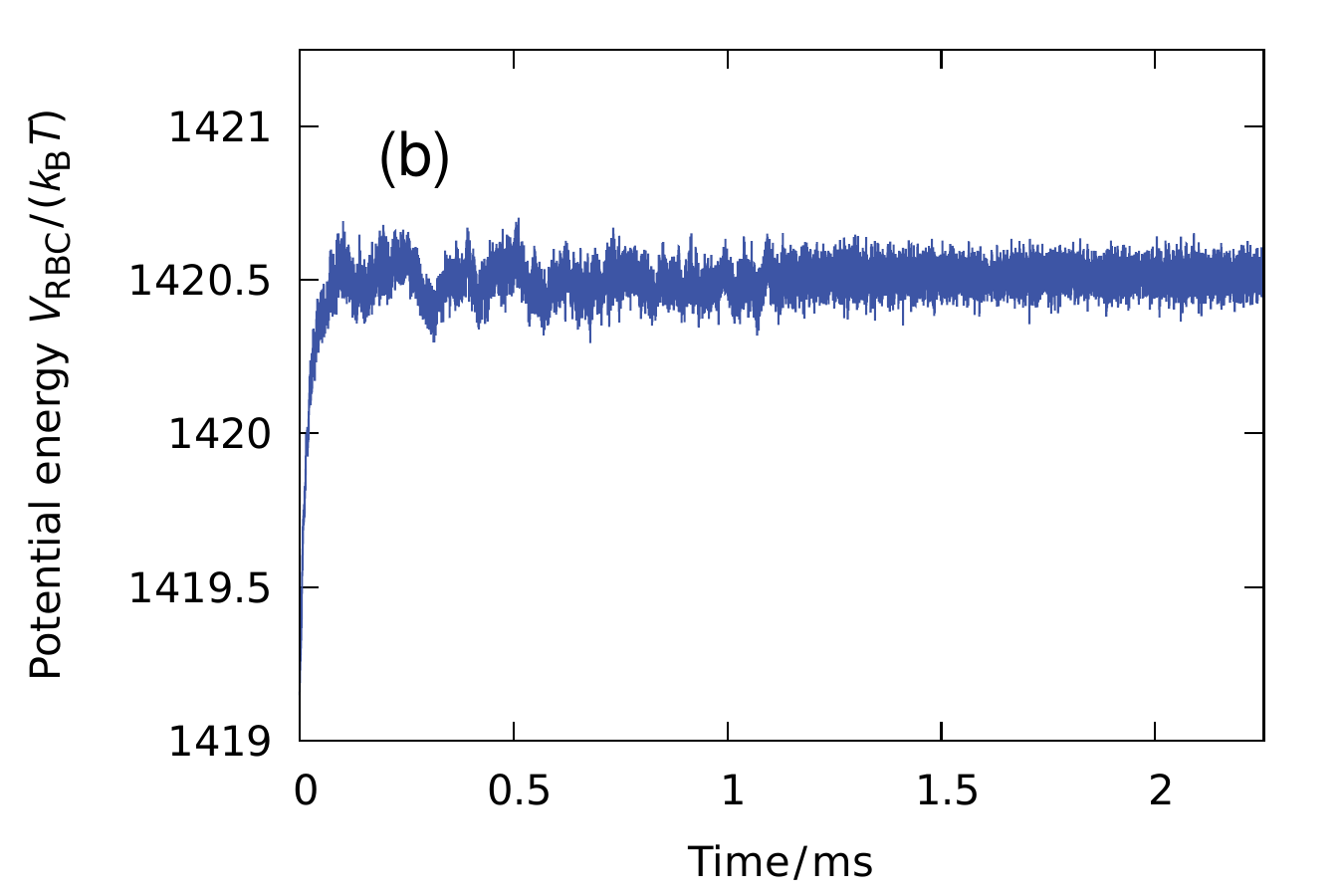}
        \label{subfig:potential}
    \end{subfigure}
    \caption{Effect of the DPD thermostat in terms of the time evolutions of (a) the total energy and (b) the total potential energy $V_\mathrm{RBC}$. The thermostat was turned off $\nunit{1.1}{ms}$ into the simulation, at which point the simulation switched to NVE. As a result, the total energy was conserved, whereas the potential energy continued to fluctuate.}
    \label{fig:dpd}
\end{figure}

The time evolutions of the average total energy and the total potential energy $V_\mathrm{RBC}$ per particle are shown in Figs.~\ref{fig:dpd} (a) and (b), respectively. We applied the DPD thermostat for the first $\nunit{1.1}{ms}$, where the relaxation time $\tau$ calculated from an exponential fit is $\nunit{0.02}{ms}$. We then continued the simulation for another $\nunit{1.1}{ms}$ with the thermostat turned off. Reflecting this condition, the total energy was conserved in the second half of the simulation. On the other hand, the potential energy continued to fluctuate.

\subsection{Spectra of potential energies} \label{subsec:spectra_potentials}

\subsubsection{Spectra and characteristic peaks} \label{subsubsec:default_spectra}

\begin{figure}[htbp]
    \begin{subfigure}[b]{.49\textwidth}
        \centering
        \includegraphics[width=\linewidth]{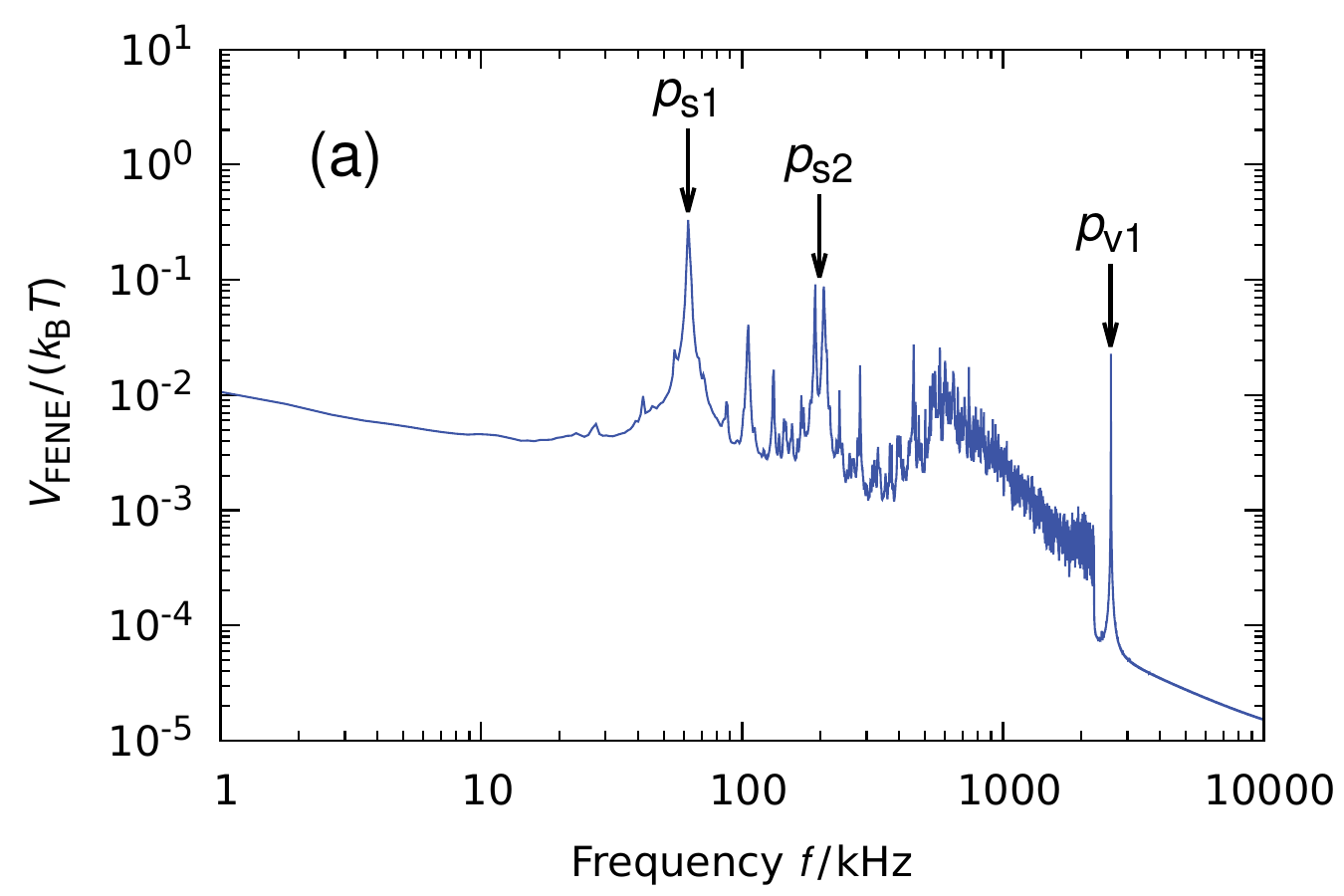}
        \label{subfig:default_spring}
    \end{subfigure}
    \hfill
    \begin{subfigure}[b]{.49\textwidth}
        \centering
        \includegraphics[width=\linewidth]{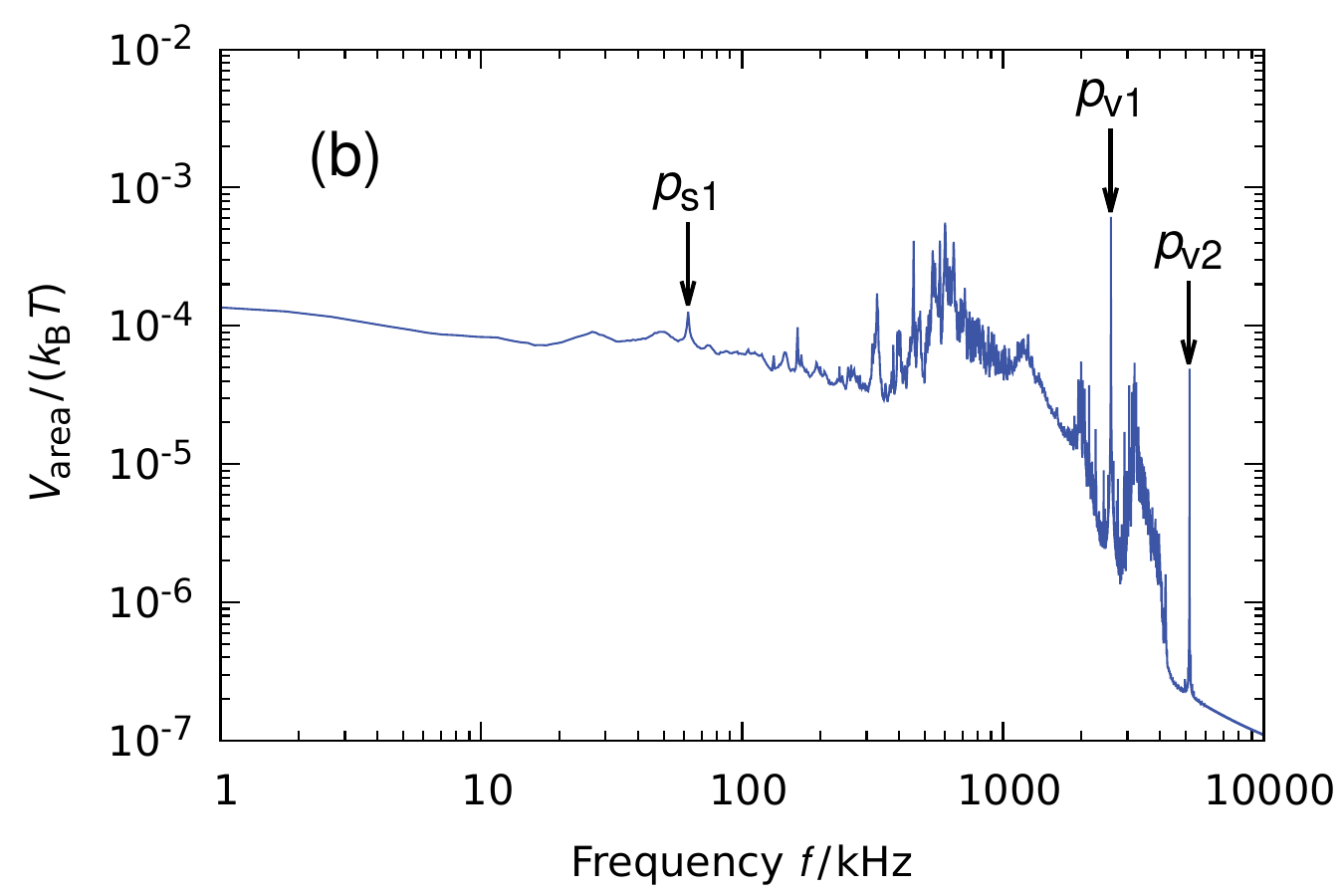}
        \label{subfig:default_area}
    \end{subfigure}
    \hfill
    \begin{subfigure}[b]{.49\textwidth}
        \centering
        \includegraphics[width=\linewidth]{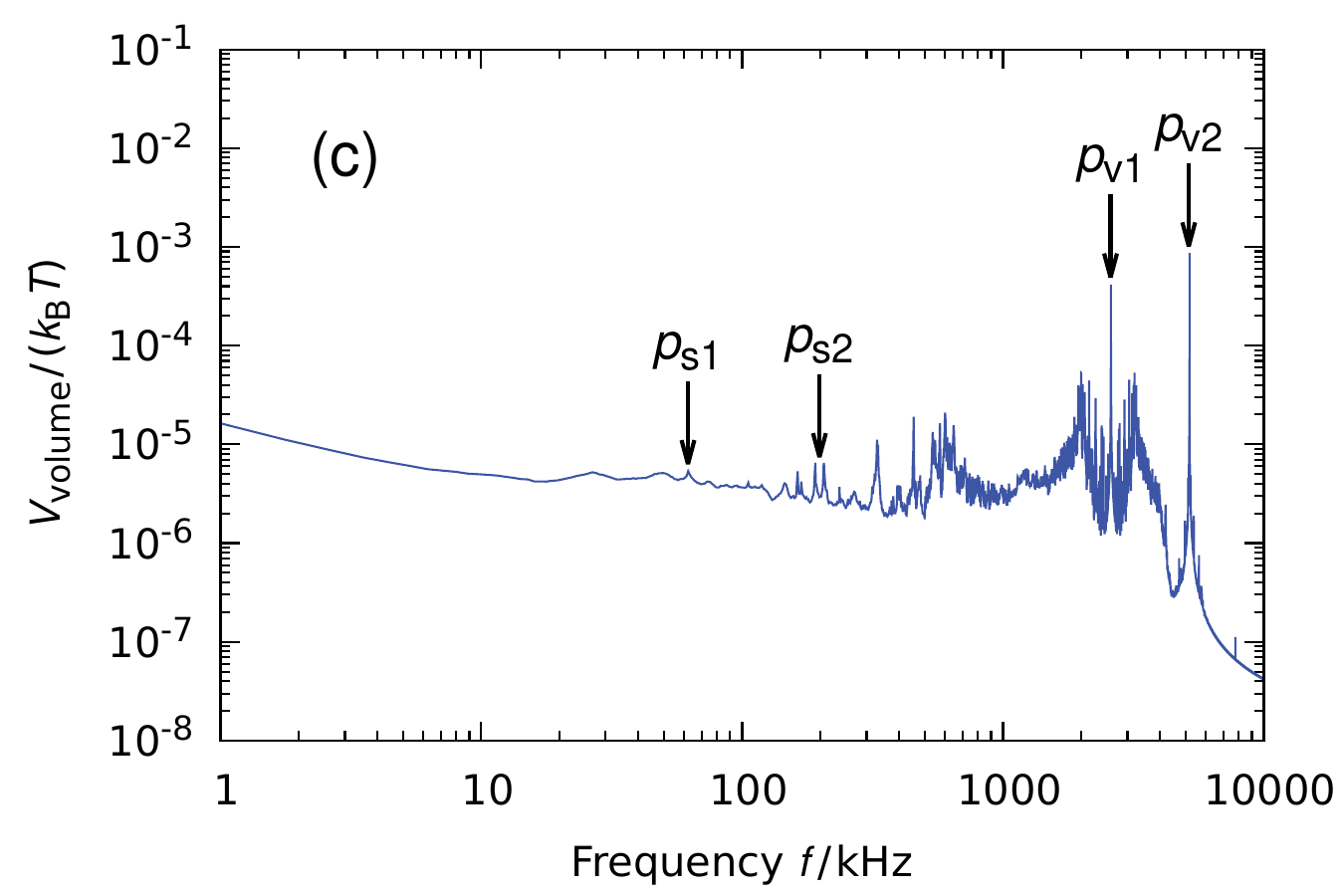}
        \label{subfig:default_volume}
    \end{subfigure}
    \hfill
    \begin{subfigure}[b]{.49\textwidth}
        \centering
        \includegraphics[width=\linewidth]{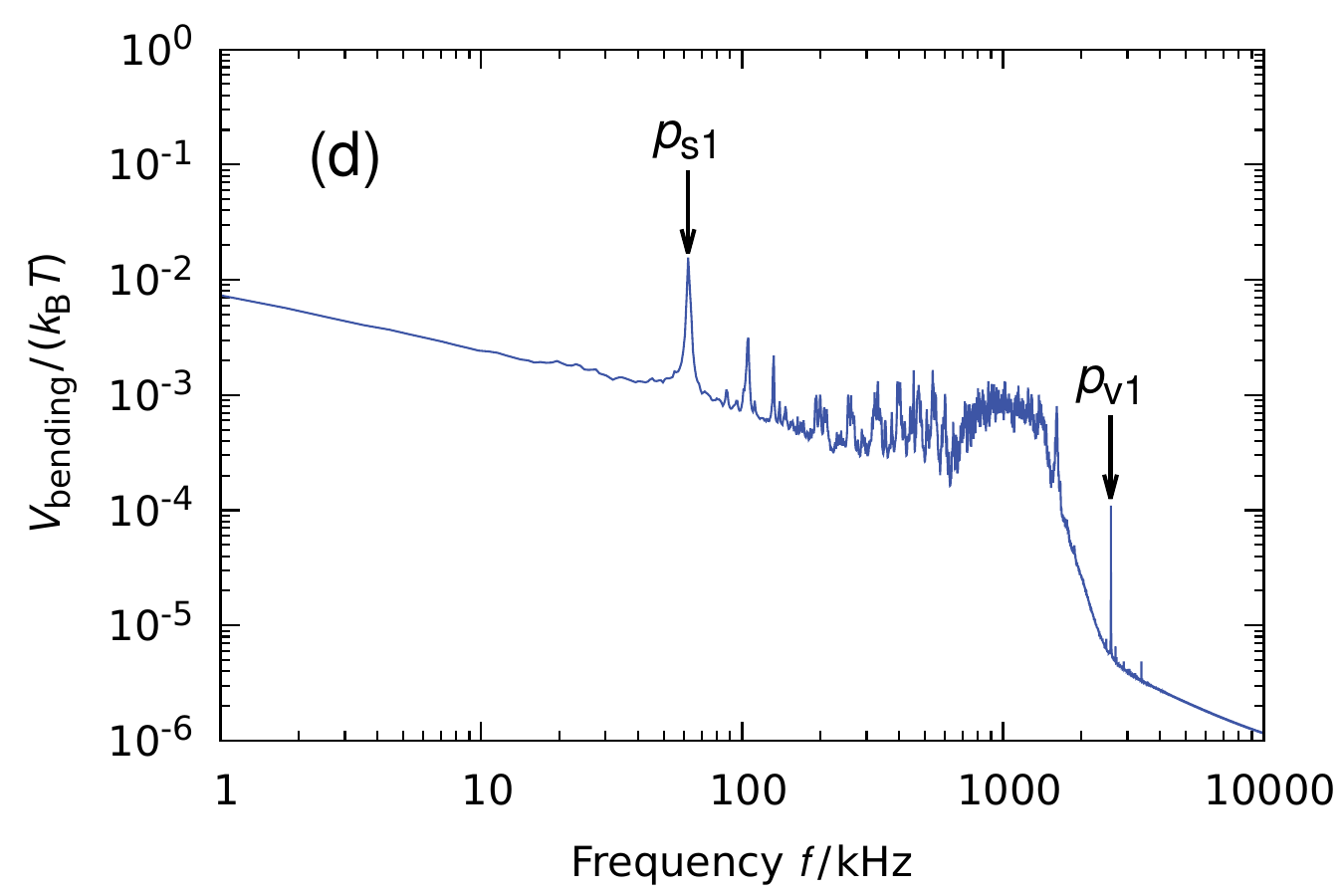}
        \label{subfig:default_bending}
    \end{subfigure}
    \caption{Fourier spectra of fluctuations of membrane potentials: (a) $V_\mathrm{FENE}$, (b) $V_\mathrm{area}$, (c) $V_\mathrm{volume}$, and (d) $V_\mathrm{bending}$. The peak $p_\mathrm{s1}$ is at $\nunit{62}{kHz}$, $p_\mathrm{s2}$ at $\nunit{200}{kHz}$, $p_\mathrm{v1}$ at $\nunit{2.6}{MHz}$, and $p_\mathrm{v2}$ at $\nunit{5.2}{MHz}$. Each peak is observed in more than one spectrum.}
    \label{fig:default_spectra}
\end{figure}

As given by Eq.~\eqref{eq:Vrbc}, the potential energy $V_\mathrm{RBC}$ is the sum of four different potentials, namely, $V_\mathrm{FENE}$, $V_\mathrm{area}$, $V_\mathrm{volume}$, and $V_\mathrm{bending}$. The Fourier spectra of these potentials are shown in Figs.~\ref{fig:default_spectra} (a), (b), (c), and (d), respectively. The results of $V_\mathrm{FENE}$ will represent the results of $V_\mathrm{POW}$ and $V_\mathrm{spring}$ because the spectrum of $V_\mathrm{POW}$ was found to be identical to that of $V_\mathrm{FENE}$, and $V_\mathrm{spring}$ is simply the sum of $V_\mathrm{FENE}$ and $V_\mathrm{POW}$. Distinct peaks appear in each spectrum, several of which are observed in many figures. Hence, we named four peaks at different frequencies as identified in the figures as follows: $p_\mathrm{s1}$ at $\nunit{62}{kHz}$, $p_\mathrm{s2}$ at $\nunit{200}{kHz}$, $p_\mathrm{v1}$ at $\nunit{2.6}{MHz}$, and $p_\mathrm{v2}$ at $\nunit{5.2}{MHz}$. In Fig.~\ref{fig:default_spectra} (a) for $V_\mathrm{FENE}$, the peaks $p_\mathrm{s1},\ p_\mathrm{s2}$, and $p_\mathrm{v1}$ appear, whereas in Fig.~\ref{fig:default_spectra} (b) for $V_\mathrm{area}$, the peaks $p_\mathrm{s1},\ p_\mathrm{v1}$, and $p_\mathrm{v2}$ are observed. All four peaks are visible in Fig.~\ref{fig:default_spectra} (c) for $V_\mathrm{volume}$, whereas the peaks $p_\mathrm{s1}$ and $p_\mathrm{v1}$ are seen in Fig.~\ref{fig:default_spectra} (d) for $V_\mathrm{bending}$.

\subsubsection{Parameter dependence of spectra}

\begin{figure}[htbp]
    \begin{subfigure}[b]{.49\textwidth}
        \centering
        \includegraphics[width=\linewidth]{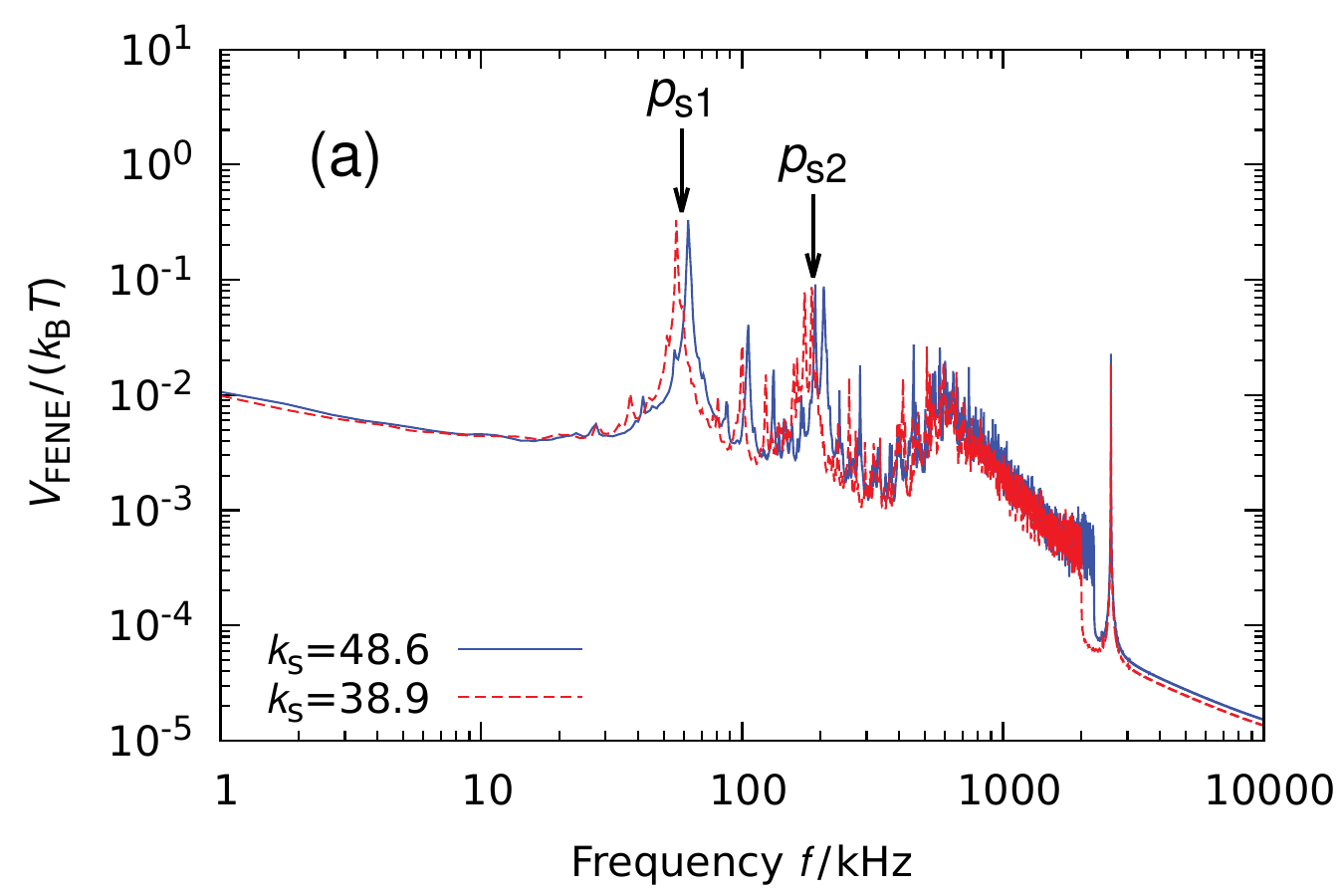}
        \label{subfig:spring}
    \end{subfigure}
    \hfill
    \begin{subfigure}[b]{.49\textwidth}
        \centering
        \includegraphics[width=\linewidth]{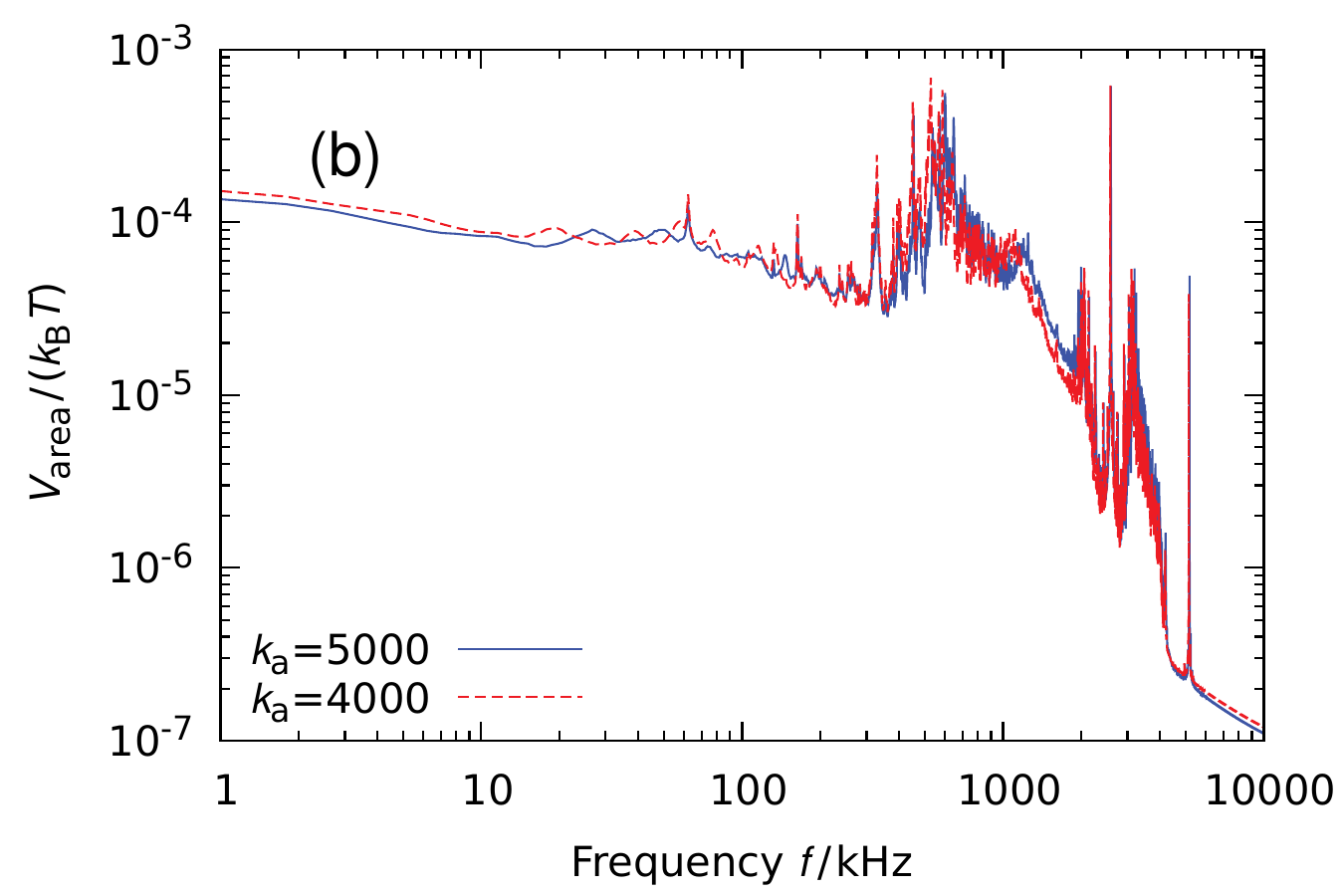}
        \label{subfig:area}
    \end{subfigure}
    \hfill
    \begin{subfigure}[b]{.49\textwidth}
        \centering
        \includegraphics[width=\linewidth]{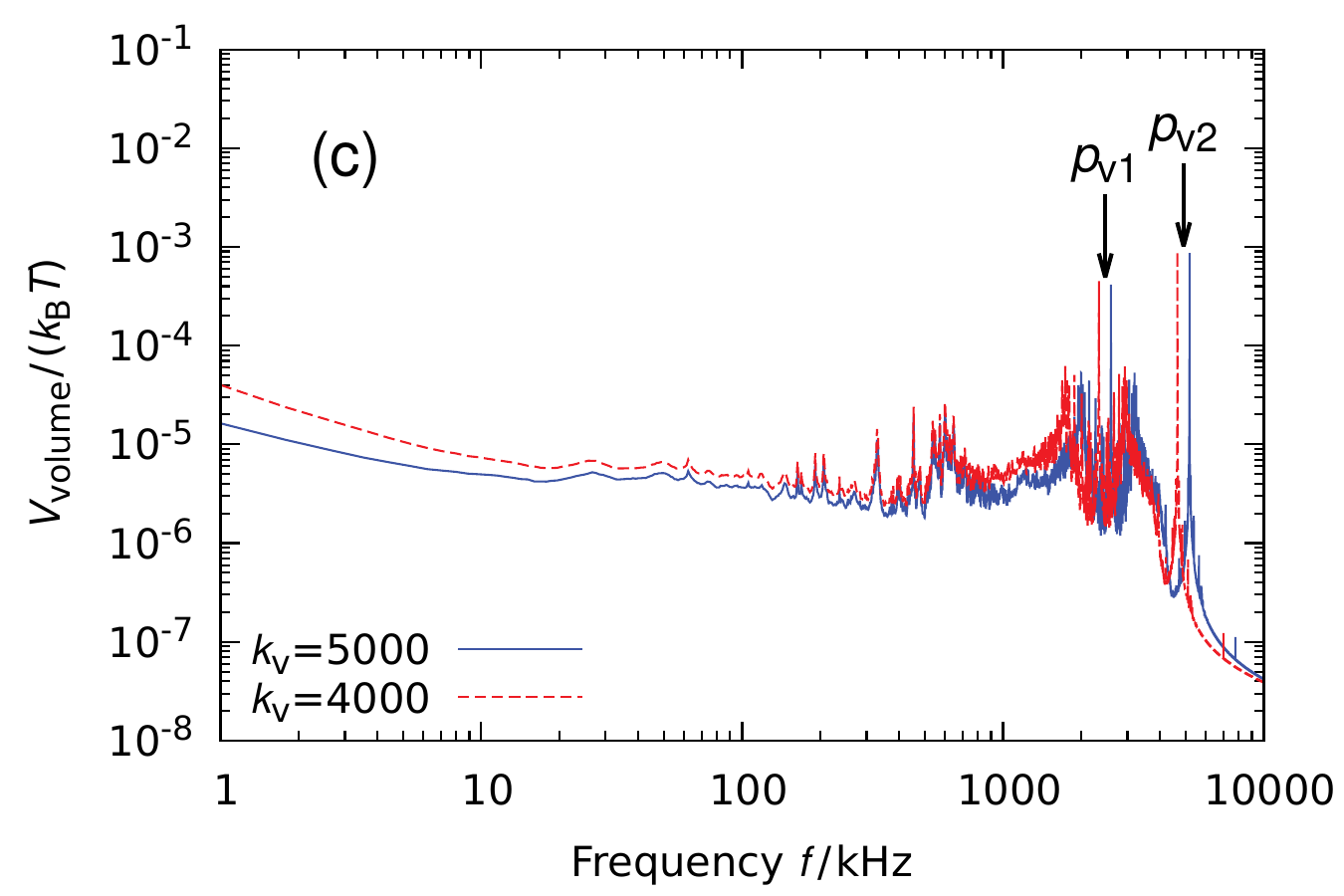}
        \label{subfig:volume}
    \end{subfigure}
    \hfill
    \begin{subfigure}[b]{.49\textwidth}
        \centering
        \includegraphics[width=\linewidth]{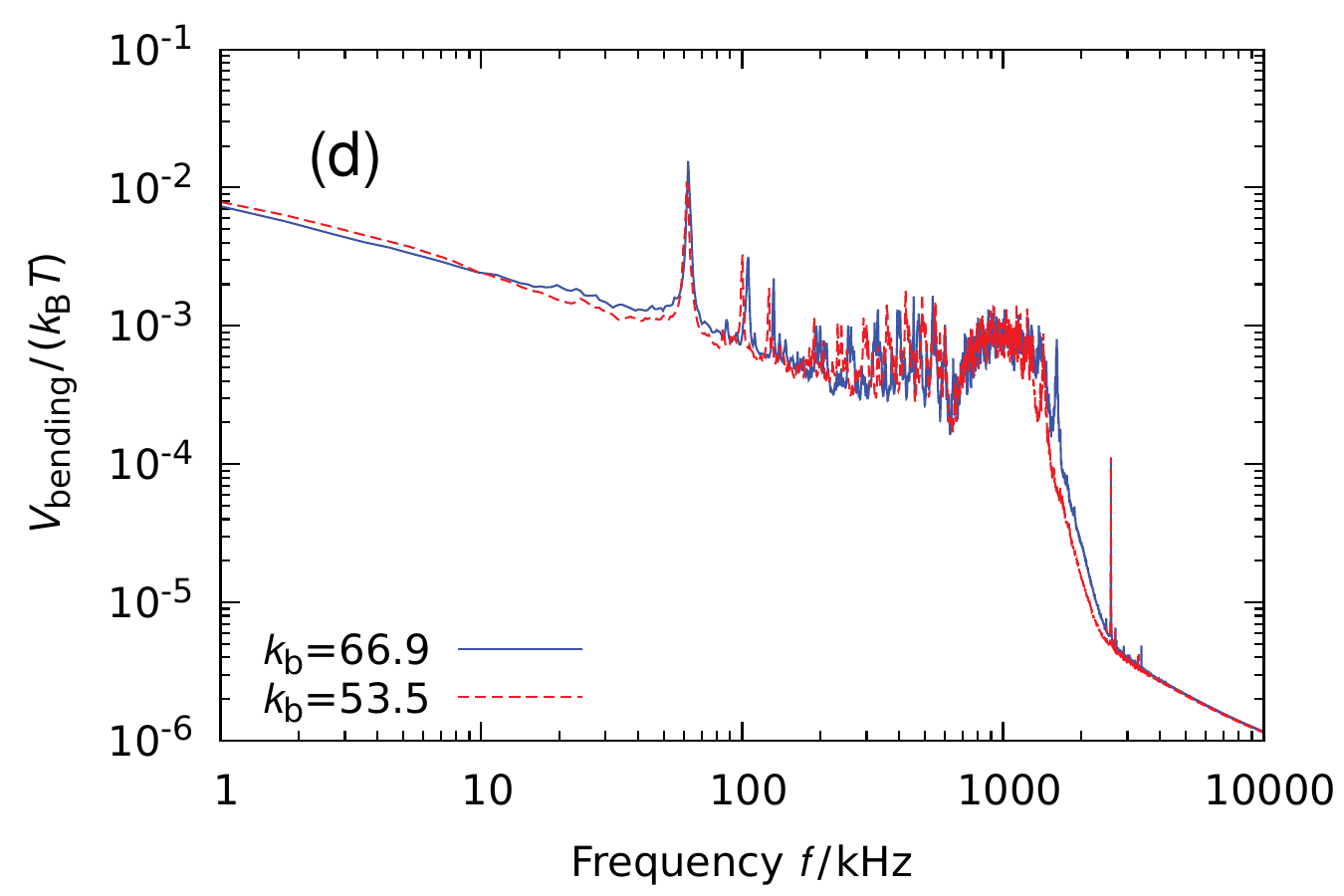}
        \label{subfig:bending}
    \end{subfigure}
    \caption{(Color online) Fourier spectra of fluctuations of membrane potentials: (a) $V_\mathrm{FENE}$, (b) $V_\mathrm{area}$, (c) $V_\mathrm{volume}$, and (d) $V_\mathrm{bending}$. The original spectra shown in Fig.~\ref{fig:default_spectra} are represented as solid lines, and the dashed lines are the new spectra where the values of the parameters $k_\mathrm{s},\ k_\mathrm{a},\ k_\mathrm{v}$, and $k_\mathrm{b}$ were altered. The peak shifts of $p_\mathrm{s1}$ and $p_\mathrm{s2}$ are observed only in (a) for $V_\mathrm{FENE}$, whereas the peak shifts of $p_\mathrm{v1}$ and $p_\mathrm{v2}$ are only seen in (c) for $V_\mathrm{volume}$.}
    \label{fig:alt_spectra}
\end{figure}

We hypothesized that the peaks observed in many spectra at identical frequencies originated from the same potential. To identify the origin of each peak, we performed another set of simulations. The values of the parameters $k_\mathrm{s}$ of $V_\mathrm{FENE}$, $k_\mathrm{a}$ of $V_\mathrm{area}$, $k_\mathrm{v}$ of $V_\mathrm{volume}$, and $k_\mathrm{b}$ of $V_\mathrm{bending}$ were reduced by 20\% from their original values. The results are shown in Fig.~\ref{fig:alt_spectra}, where the newly obtained spectra (dashed line) are superimposed on the original spectra of Fig.~\ref{fig:default_spectra} (solid line). As in Fig.~\ref{fig:default_spectra}, the spectra of $V_\mathrm{FENE},\ V_\mathrm{area},\ V_\mathrm{volume}$, and $V_\mathrm{bending}$ are labeled (a), (b), (c), and (d), respectively. Altering the values of the parameters resulted in an overall shift in all the spectra. However, the peaks $p_\mathrm{s1}$ and $p_\mathrm{s2}$ have shifted only in Fig.~\ref{fig:alt_spectra} (a) for $V_\mathrm{FENE}$, whereas Fig.~\ref{fig:alt_spectra} (c) for $V_\mathrm{volume}$ is the only figure in which $p_\mathrm{v1}$ and $p_\mathrm{v2}$ are seen to shift. Moreover, no distinct peak shifts are observed in Fig.~\ref{fig:alt_spectra} (b) for $V_\mathrm{area}$ and Fig.~\ref{fig:alt_spectra} (d) for $V_\mathrm{bending}$. This implies that all the characteristic peaks identifiable in the Fourier spectra of membrane potentials originate from $V_\mathrm{FENE}$ and $V_\mathrm{volume}$.

\subsection{Spectra of the radius of gyration}

\begin{figure}[htbp]
    \begin{subfigure}[b]{.49\textwidth}
        \centering
        \includegraphics[width=\linewidth]{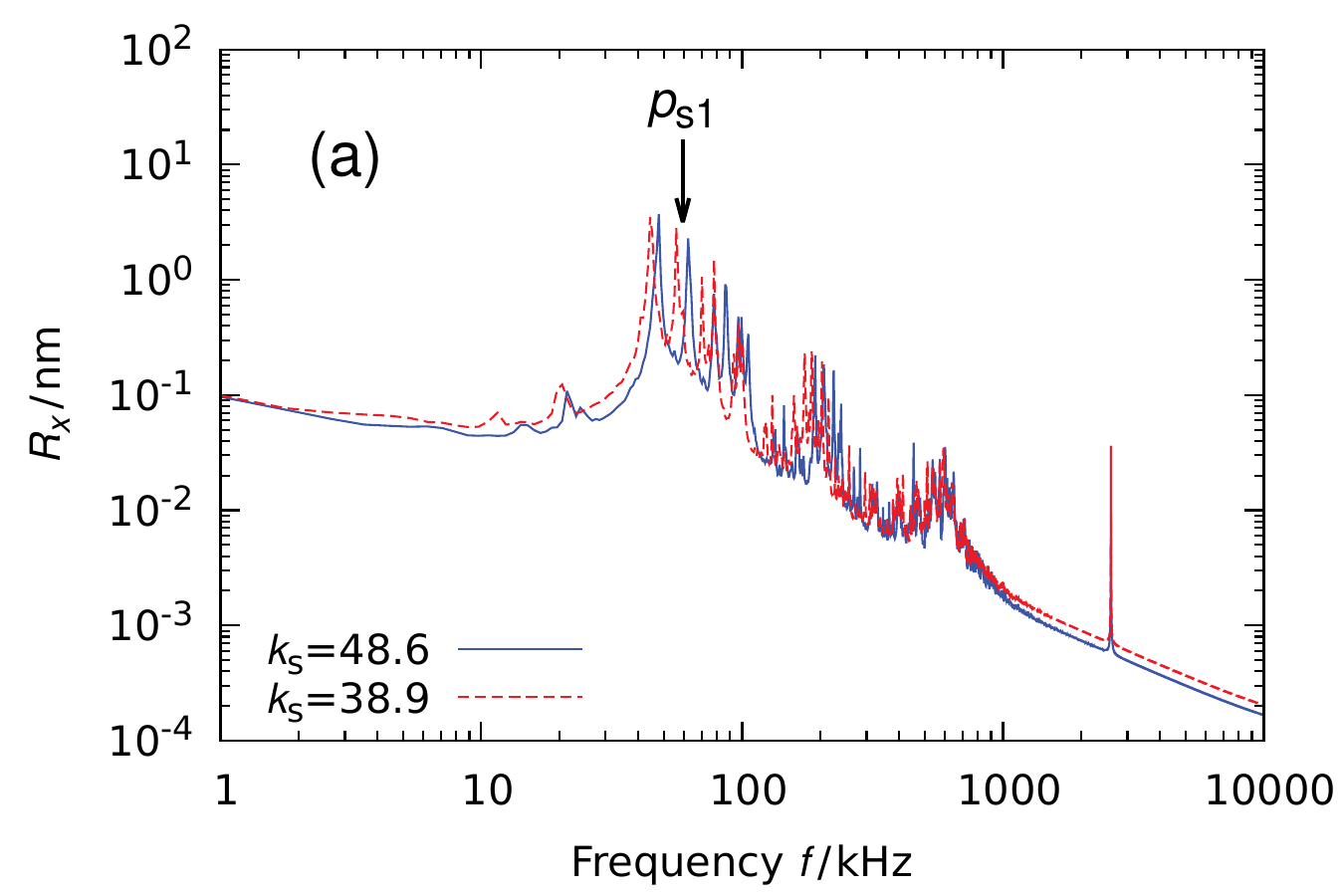}
        \label{subfig:rx_spring}
    \end{subfigure}
    \hfill
    \begin{subfigure}[b]{.49\textwidth}
        \centering
        \includegraphics[width=\linewidth]{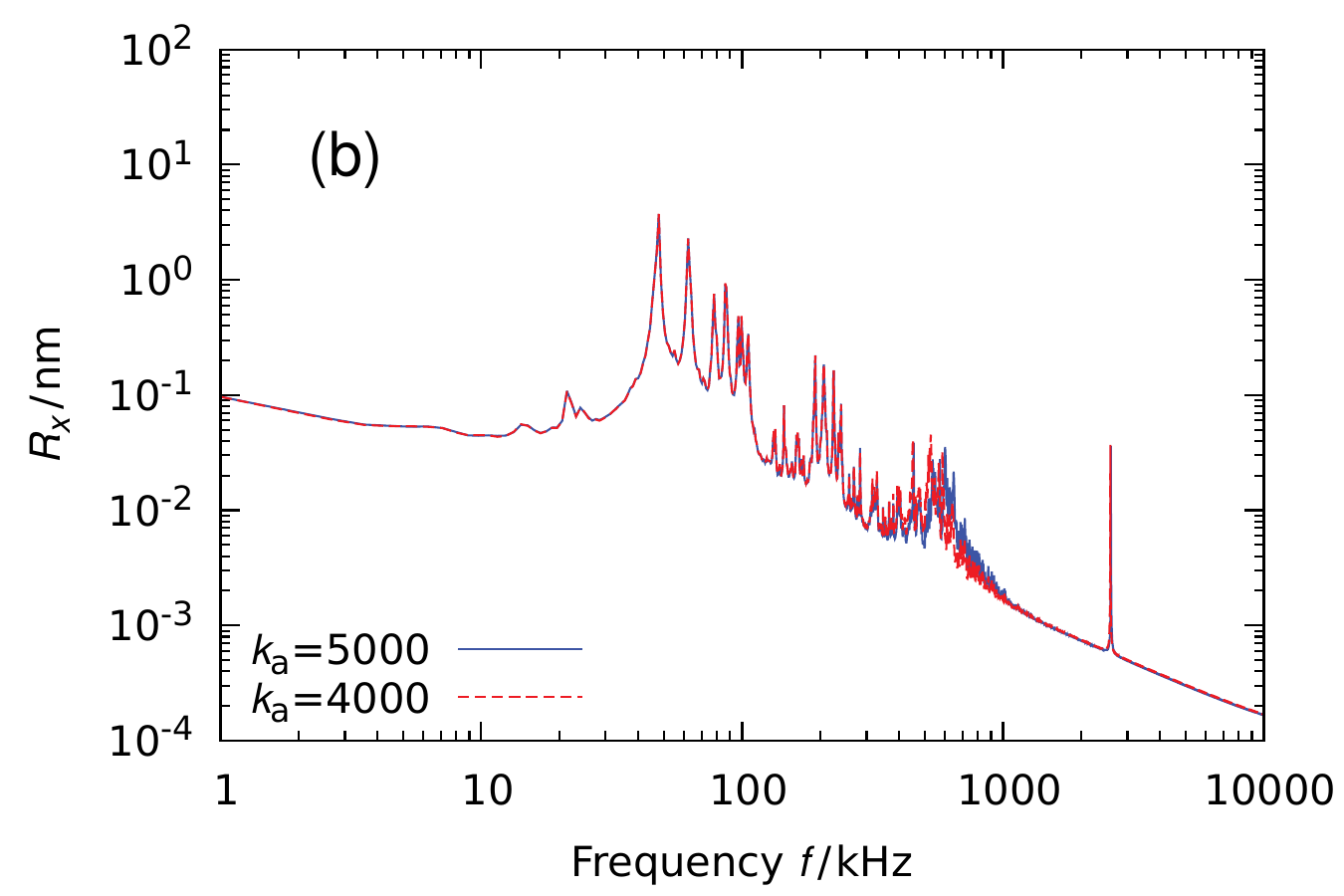}
        \label{subfig:rx_area}
    \end{subfigure}
    \hfill
    \begin{subfigure}[b]{.49\textwidth}
        \centering
        \includegraphics[width=\linewidth]{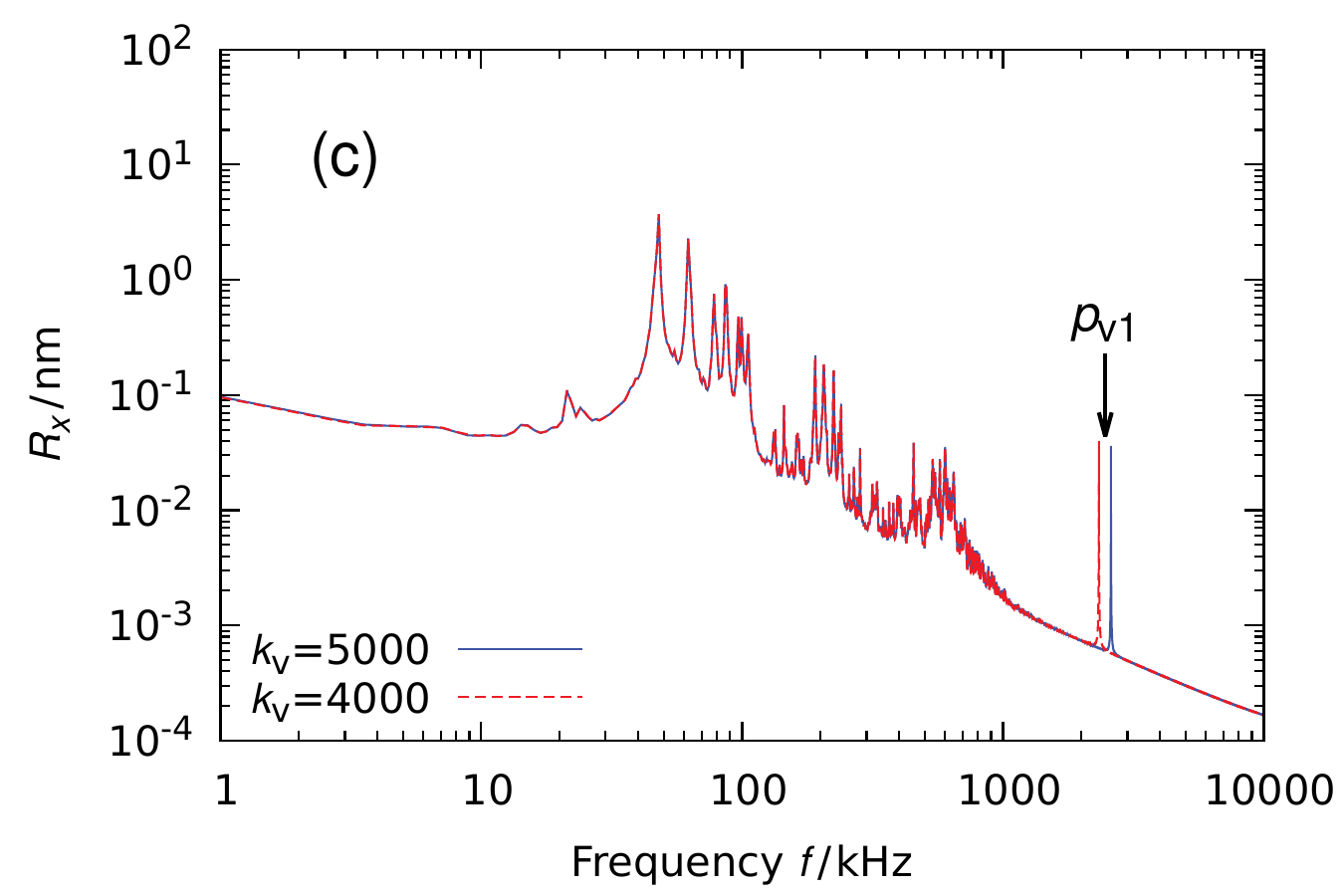}
        \label{subfig:rx_volume}
    \end{subfigure}
    \hfill
    \begin{subfigure}[b]{.49\textwidth}
        \centering
        \includegraphics[width=\linewidth]{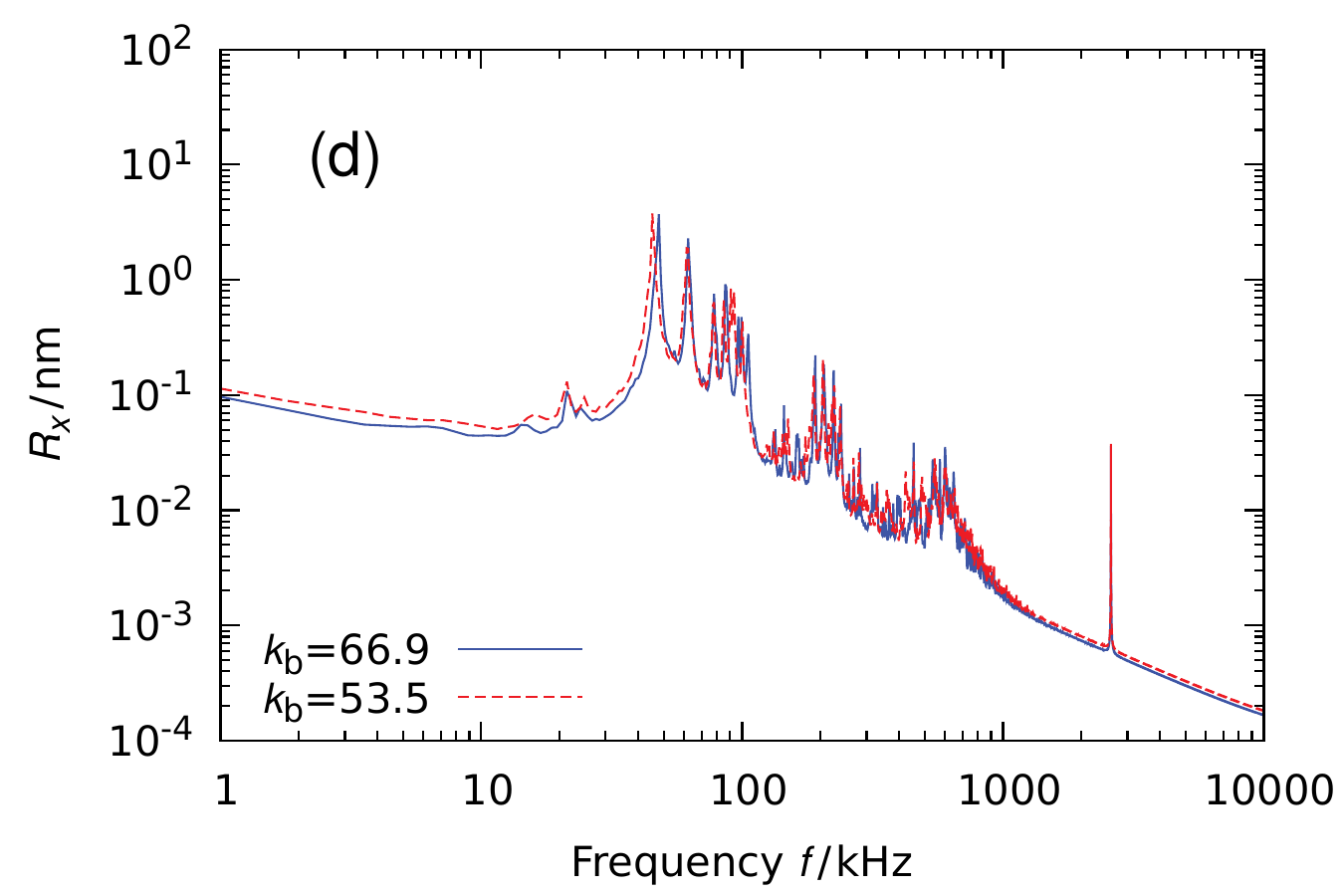}
        \label{subfig:rx_bending}
    \end{subfigure}
    \caption{(Color online) Fourier spectra of the radius of gyration $R_x$. The parameters $k_\mathrm{s},\ k_\mathrm{a},\ k_\mathrm{v}$, and $k_\mathrm{b}$ were each altered in their corresponding figures in (a), (b), (c), and (d). The peak $p_\mathrm{s1}$ has a frequency of $\nunit{62}{kHz}$ and $p_\mathrm{v1}$ a frequency of $\nunit{2.6}{MHz}$.}
    \label{fig:rx_spectra}
\end{figure}

\begin{figure}[htbp]
    \begin{subfigure}[b]{.49\textwidth}
        \centering
        \includegraphics[width=\linewidth]{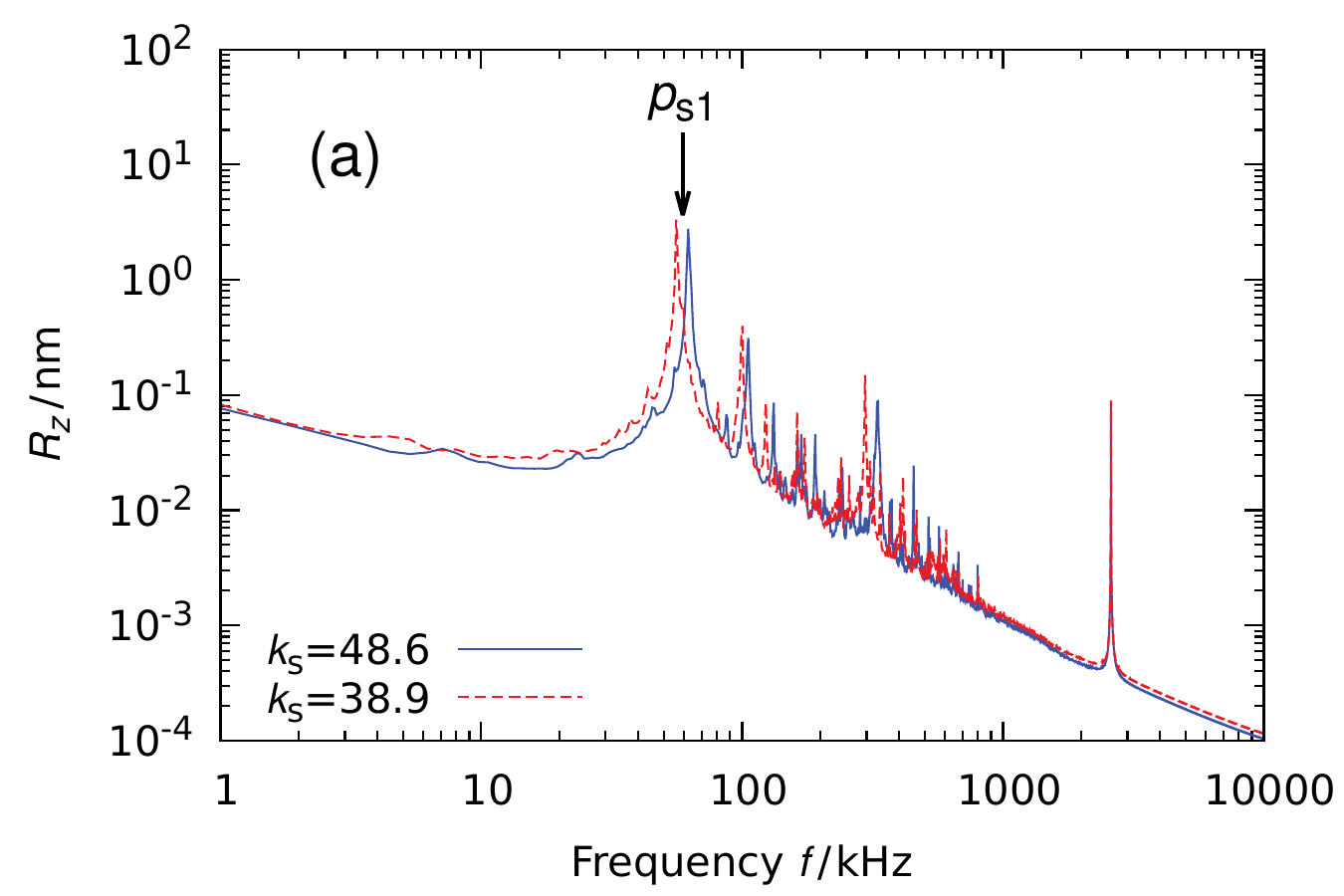}
        \label{subfig:rz_spring}
    \end{subfigure}
    \hfill
    \begin{subfigure}[b]{.49\textwidth}
        \centering
        \includegraphics[width=\linewidth]{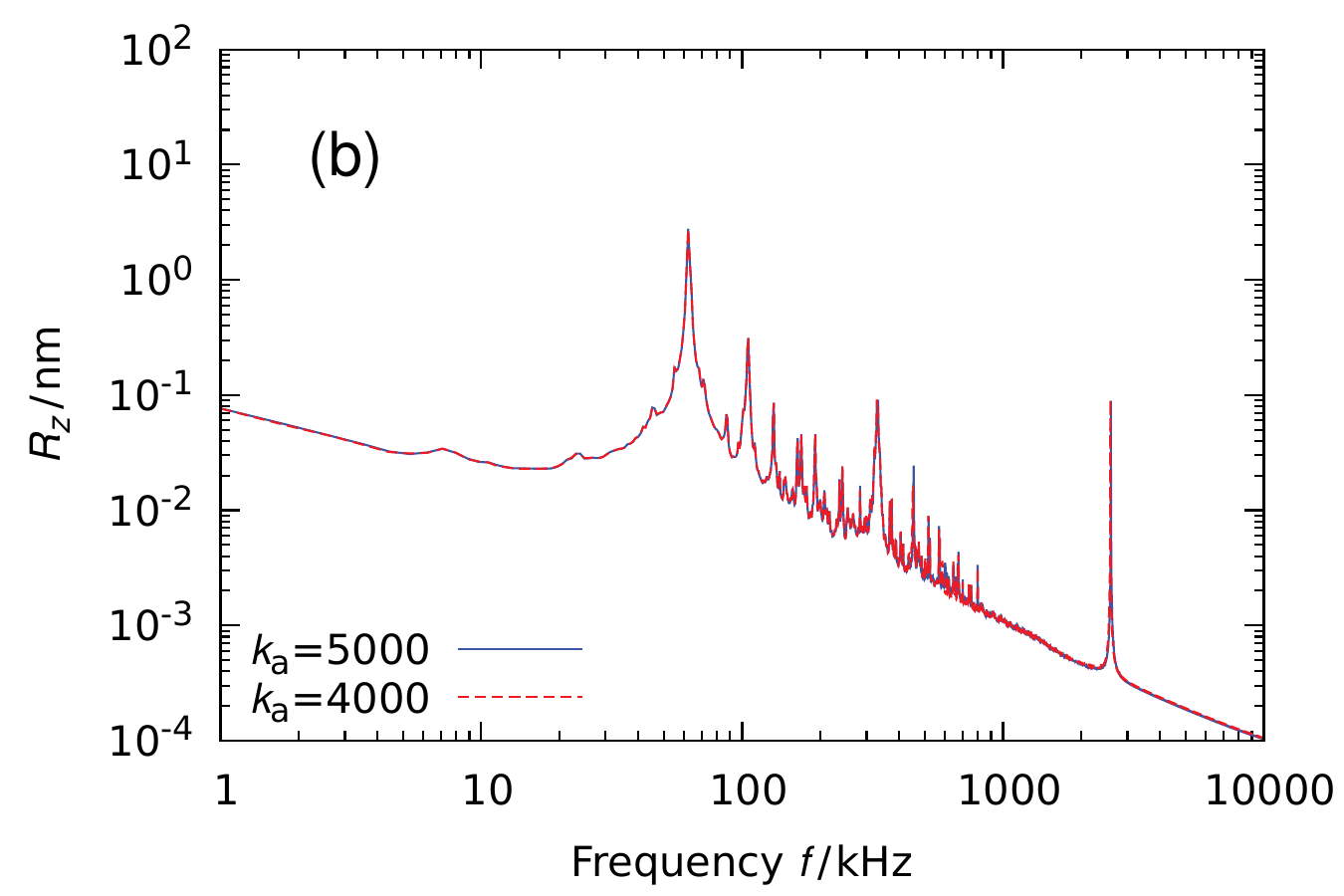}
        \label{subfig:rz_area}
    \end{subfigure}
    \hfill
    \begin{subfigure}[b]{.49\textwidth}
        \centering
        \includegraphics[width=\linewidth]{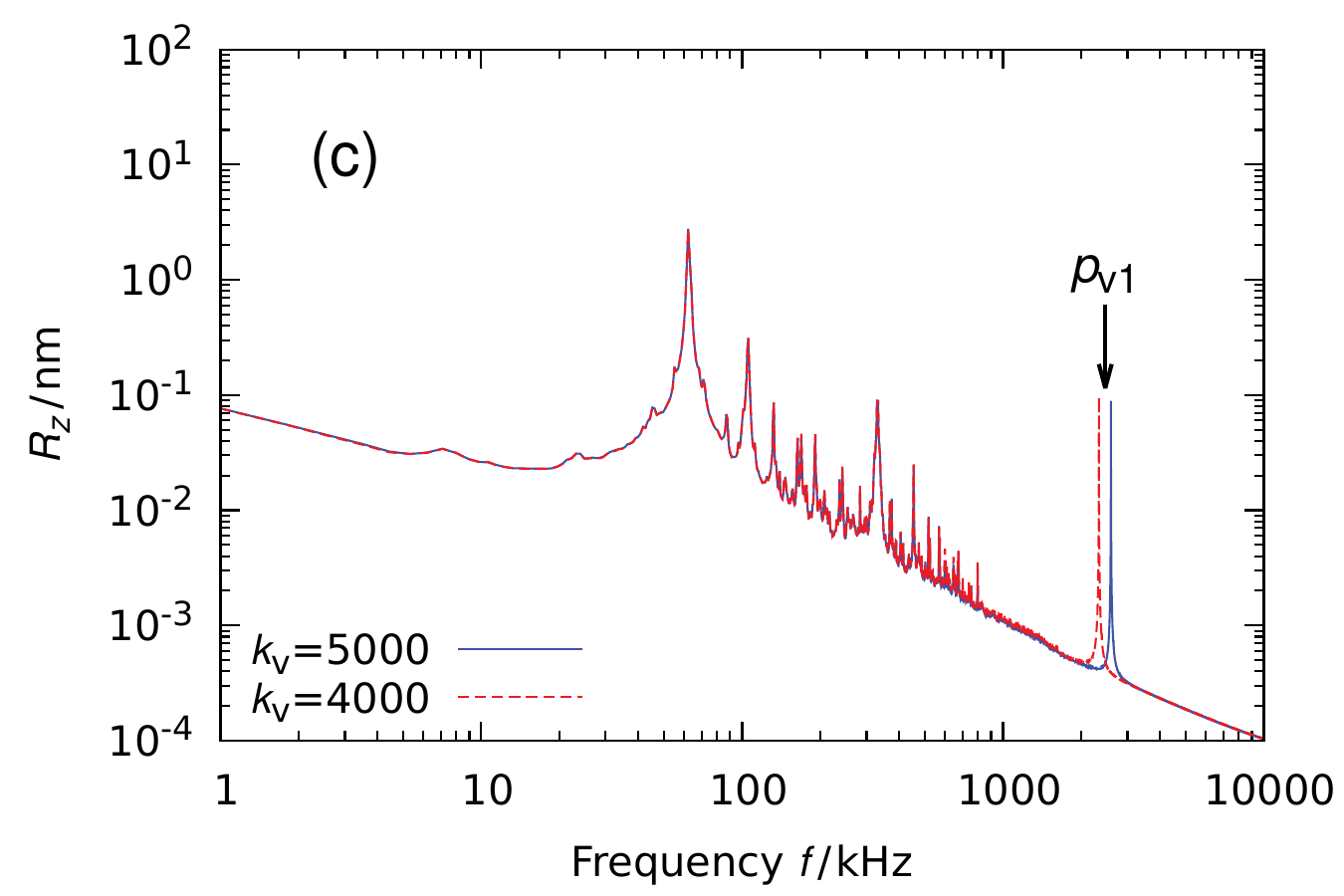}
        \label{subfig:rz_volume}
    \end{subfigure}
    \hfill
    \begin{subfigure}[b]{.49\textwidth}
        \centering
        \includegraphics[width=\linewidth]{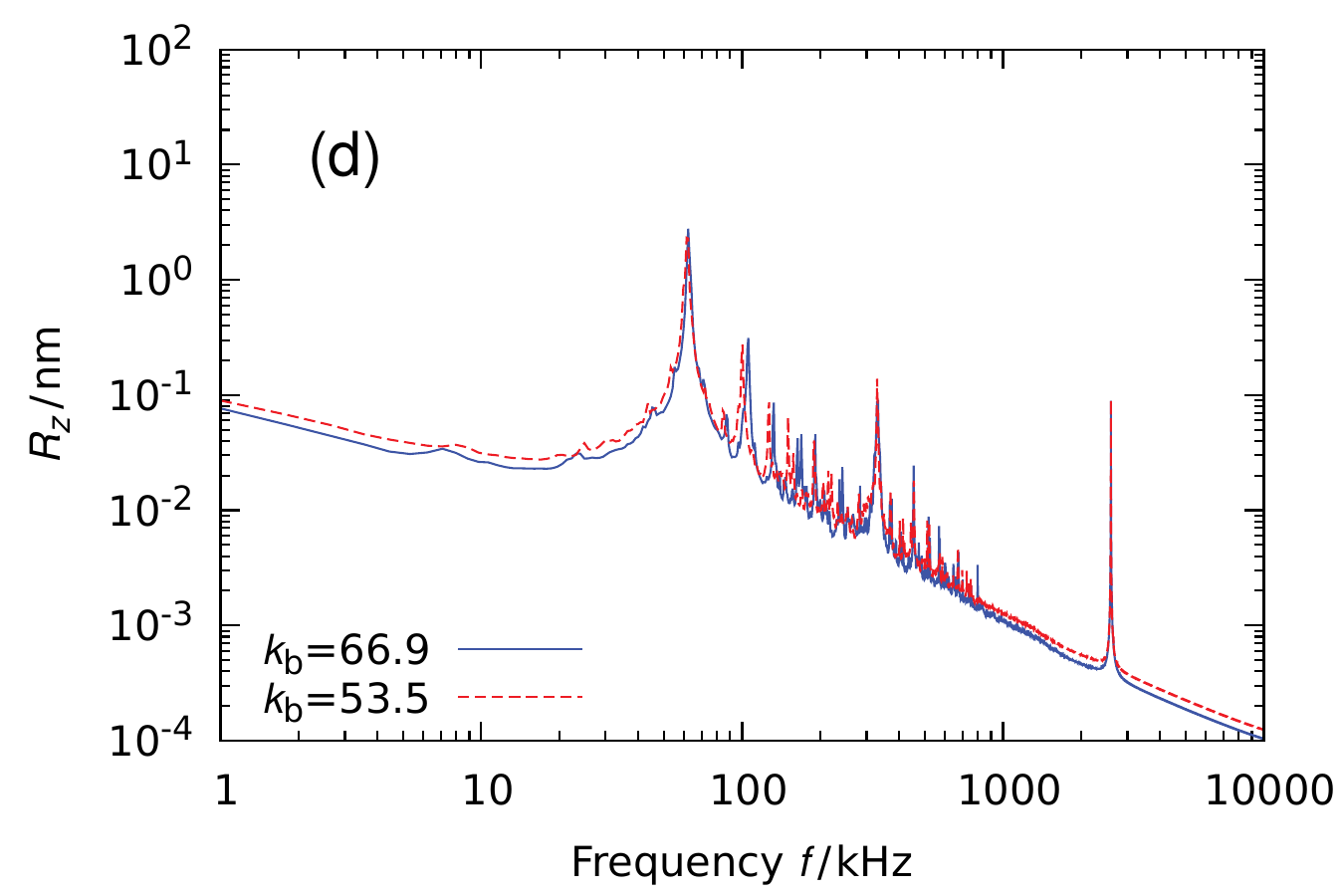}
        \label{subfig:rz_bending}
    \end{subfigure}
    \caption{(Color online) Fourier spectra of the radius of gyration $R_z$. The parameters $k_\mathrm{s},\ k_\mathrm{a},\ k_\mathrm{v}$, and $k_\mathrm{b}$ were each altered in their corresponding figures in (a), (b), (c), and (d). The peak $p_\mathrm{s1}$ has a frequency of $\nunit{62}{kHz}$ and $p_\mathrm{v1}$ a frequency of $\nunit{2.6}{MHz}$.}
    \label{fig:rz_spectra}
\end{figure}

Although we have thus far identified the peaks observed in the Fourier spectra of the membrane potentials, the potentials themselves are not directly measurable by experiments. Accordingly, we measured the radii of gyration $R_x,\ R_y$, and $R_z$ given by Eq.~\eqref{eq:gyr_radius}. Considering the rotational symmetry of the RBC in the $xy$-plane, the spectra of $R_x$ are shown in Fig.~\ref{fig:rx_spectra}, whereas the spectra of $R_z$ are shown in Fig.~\ref{fig:rz_spectra}. In both figures, the parameters $k_\mathrm{s}$ of $V_\mathrm{FENE}$, $k_\mathrm{a}$ of $V_\mathrm{area}$, $k_\mathrm{v}$ of $V_\mathrm{volume}$, and $k_\mathrm{b}$ of $V_\mathrm{bending}$ are altered in (a), (b), (c), and (d), respectively. The peak $p_\mathrm{s1}$ has a frequency of $\nunit{62}{kHz}$ and $p_\mathrm{v1}$ has a frequency of $\nunit{2.6}{MHz}$, identical to the peaks in Figs.~\ref{fig:default_spectra} and \ref{fig:alt_spectra}. In both the spectra of $R_x$ and $R_z$, no peak shifts are observed in (b) and (d), where the corresponding parameters of $V_\mathrm{area}$ and $V_\mathrm{bending}$ were altered. On the other hand, the dependence of $p_\mathrm{s1}$ on $V_\mathrm{FENE}$ is seen in (a), as is the dependence of $p_\mathrm{v1}$ on $V_\mathrm{volume}$ in (c). Although Figs.~\ref{fig:rx_spectra} and \ref{fig:rz_spectra} show similar results, an important distinction should be made regarding $p_\mathrm{s1}$. The peak is surrounded by other similar peaks in Fig.~\ref{fig:rx_spectra}, whereas it is independent of other peaks in Fig.~\ref{fig:rz_spectra}. This implies that $p_\mathrm{s1}$ is more easily observed in the spectrum of $R_z$.

\subsection{Feasibility in experiments} \label{subsec:experiments}

Experiments have been conducted where they measured the Fourier spectra of the membrane fluctuations of an RBC\cite{humpert2003,gogler2007}. However, the results of these experiments cannot be directly compared with our results due to several differences in the environment and methods of measurement. First, in experiments, the RBCs are placed in a saline solution, whereas our simulations are performed in a vacuum. Second, the observable in experiments is the position of the rim of the membrane or the attached beads, whereas the radius of gyration is measured in our simulations. Third, only the frequencies up to approximately $\nunit{100}{Hz}$ were measured in the previously mentioned experiments, which is below the frequency range of the peaks intrinsic to the RBC membrane as measured in our simulations. Therefore, in the following, we discuss the feasibility of our simulations in experiments regarding these three problems.

\subsubsection{Viscosity of the surrounding fluids} \label{subsubsec:viscosity}

To consider the effect of the surrounding fluids, we performed a set of simulations using the Langevin thermostat instead of the DPD thermostat to implicitly simulate the interactions between the membrane and the surrounding fluids. The Langevin thermostat is reflected in the equation of motion for a single particle, which is expressed as
\begin{equation}
    \label{eq:langevin}
    m\dot{\boldsymbol{v}}=\boldsymbol{F}^\text{C}-\lambda\boldsymbol{v}+\sqrt{2\lambda k_\mathrm{B}T}\boldsymbol{R}.
\end{equation}
Here, $\boldsymbol{F}^\text{C}$ is the conservative force, $\lambda$ is the damping coefficient, and the force $\boldsymbol{R}$ is a Gaussian white noise satisfying
\begin{align}
    \Braket{\boldsymbol{R}(t)}                        & =\boldsymbol{0}, \\
    \Braket{\boldsymbol{R}(t)\cdot\boldsymbol{R}(t')} & =\delta(t-t').
\end{align}
The strength of the thermostat and its effect on the membrane fluctuations are determined by $\lambda$ given a constant temperature $T$. However, we are currently unaware of the value of $\lambda$ for the Langevin thermostat corresponding to the cytoplasm and the suspending fluid in an experimental setting. Therefore, we performed simulations for different values of $\lambda$ and investigated the effect of $\lambda$ on the Fourier spectra. Considering a realistic sample size for experiments, the spectra were averaged over 100 samples for each value of $\lambda$.

\begin{figure}[htbp]
    \centering
    \includegraphics[width=.7\textwidth]{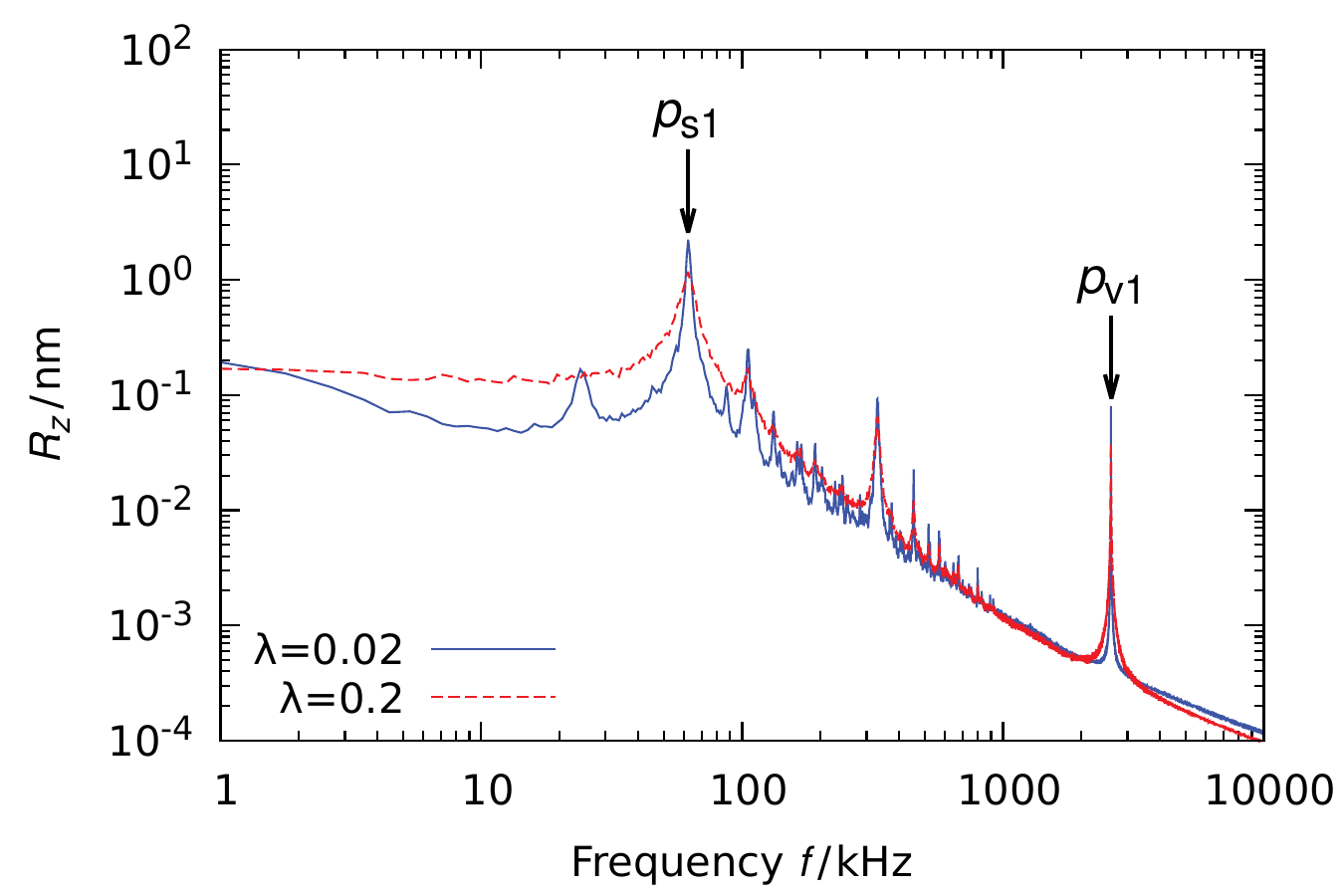}
    \caption{(Color online) Fourier spectra of the radius of gyration $R_z$ for $\lambda=0.02\text{ and }0.2$, where $\lambda$ is the damping coefficient introduced in Eq.\ \eqref{eq:langevin}. The peak $p_\mathrm{s1}$ has a frequency of $\nunit{62}{kHz}$ and $p_\mathrm{v1}$ a frequency of $\nunit{2.6}{MHz}$.}
    \label{fig:alt_gamma}
\end{figure}

Figure \ref{fig:alt_gamma} shows the spectra of the radius of gyration $R_z$ for different values of the damping coefficient $\lambda$ introduced in Eq.\ \eqref{eq:langevin}. The peaks $p_\mathrm{s1}$ and $p_\mathrm{v1}$ are detectable for both $\lambda=0.02\text{ and }0.2$, meaning that a sample size of 100 is sufficient to detect the larger peaks. The two spectra are different in that the spectrum for the larger value of $\lambda=0.2$ is smoother with broader peaks. This implies that the peaks will be less pronounced as the viscosity of the surrounding fluids increases. Because we currently do not know the physical value of $\lambda$, the peaks may not be experimentally observable due to considerable thermal noise. In that case, alternative measurements must be conducted, such as by taking the correlation of fluctuations at two points to cancel out the noise.

\subsubsection{Comparison of point fluctuations and the radius of gyration}

\begin{figure}[htbp]
    \begin{subfigure}[c]{.48\textwidth}
        \centering
        \includegraphics[width=\linewidth]{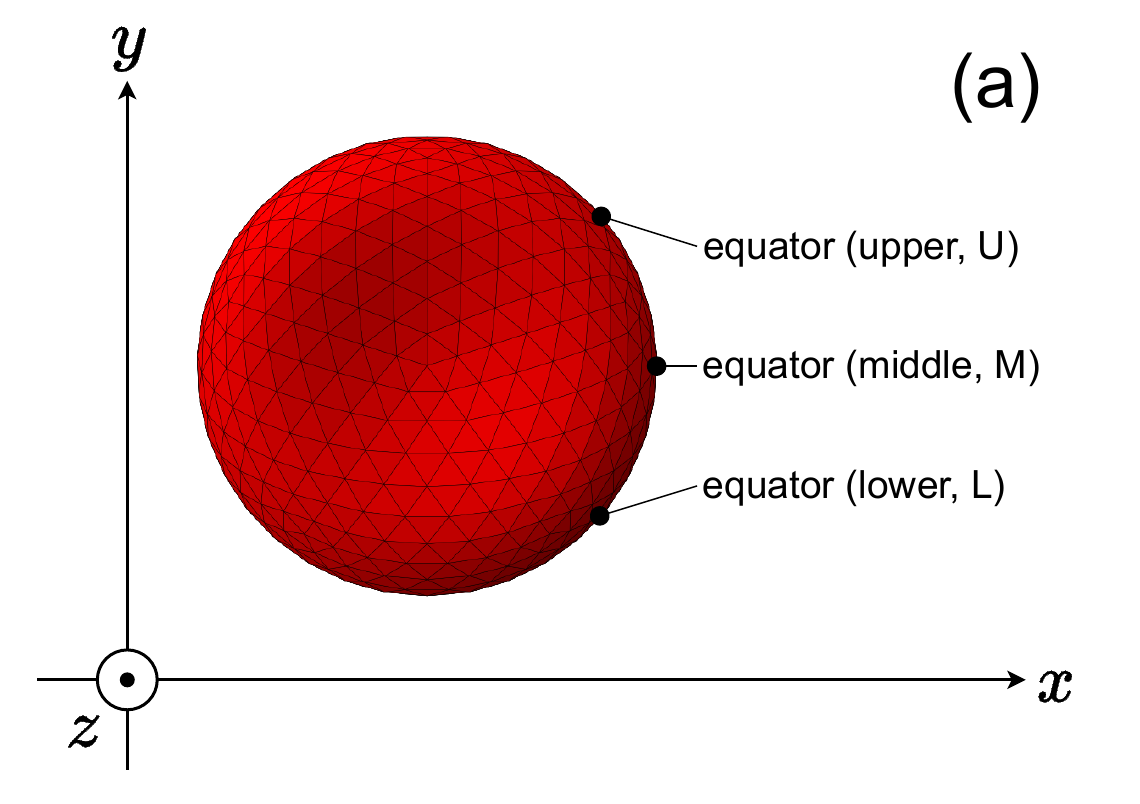}
        \label{subfig:x_points}
    \end{subfigure}
    \hfill
    \begin{subfigure}[c]{.51\textwidth}
        \centering
        \includegraphics[width=\linewidth]{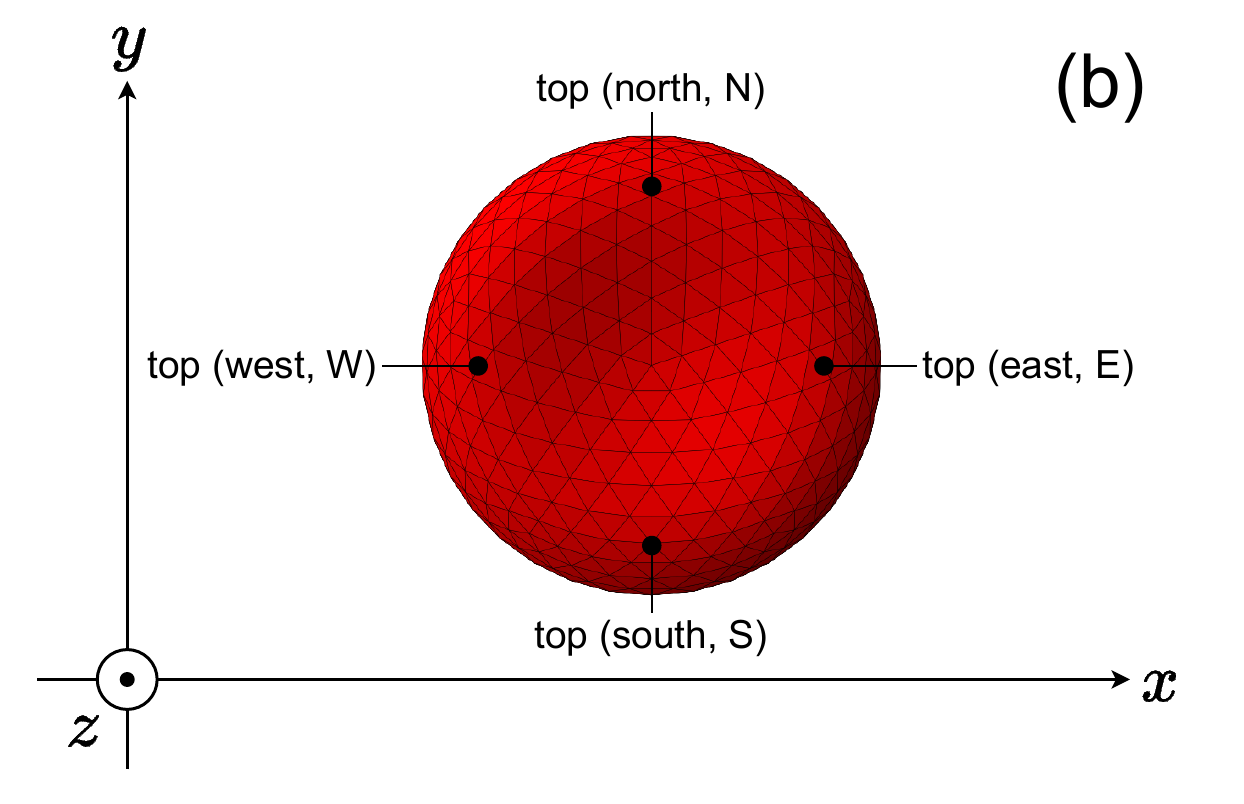}
        \label{subfig:z_points}
    \end{subfigure}
    \caption{(Color online) Single points where the membrane fluctuations are measured. Measurement in the (a) $x$-axis and (b) $z$-axis directions.}
    \label{fig:single_points}
\end{figure}

We measured the fluctuations of single points on the membrane in addition to the radius of gyration to investigate the differences between the observables. The locations of the observed points are shown in Fig.\ \ref{fig:single_points}, each composed of 7 membrane particles.
We chose three locations at the middle (M), upper (U), and lower (L) points along the equator as shown in Fig.\ \ref{fig:single_points}~(a). We chose another four locations at the east (E), west (W), north (N), and south (S) points at the top as shown in Fig.\ \ref{fig:single_points}~(b). Points on the equator were measured for the fluctuations in the $x$-axis direction, whereas those at the top were measured for the fluctuations in the $z$-axis direction. We compared the resulting spectra with those of the radii of gyration $R_x$ and $R_z$. The sample size was kept at 100 for the measurements.

\begin{figure}[htbp]
    \begin{subfigure}[b]{.49\textwidth}
        \centering
        \includegraphics[width=\linewidth]{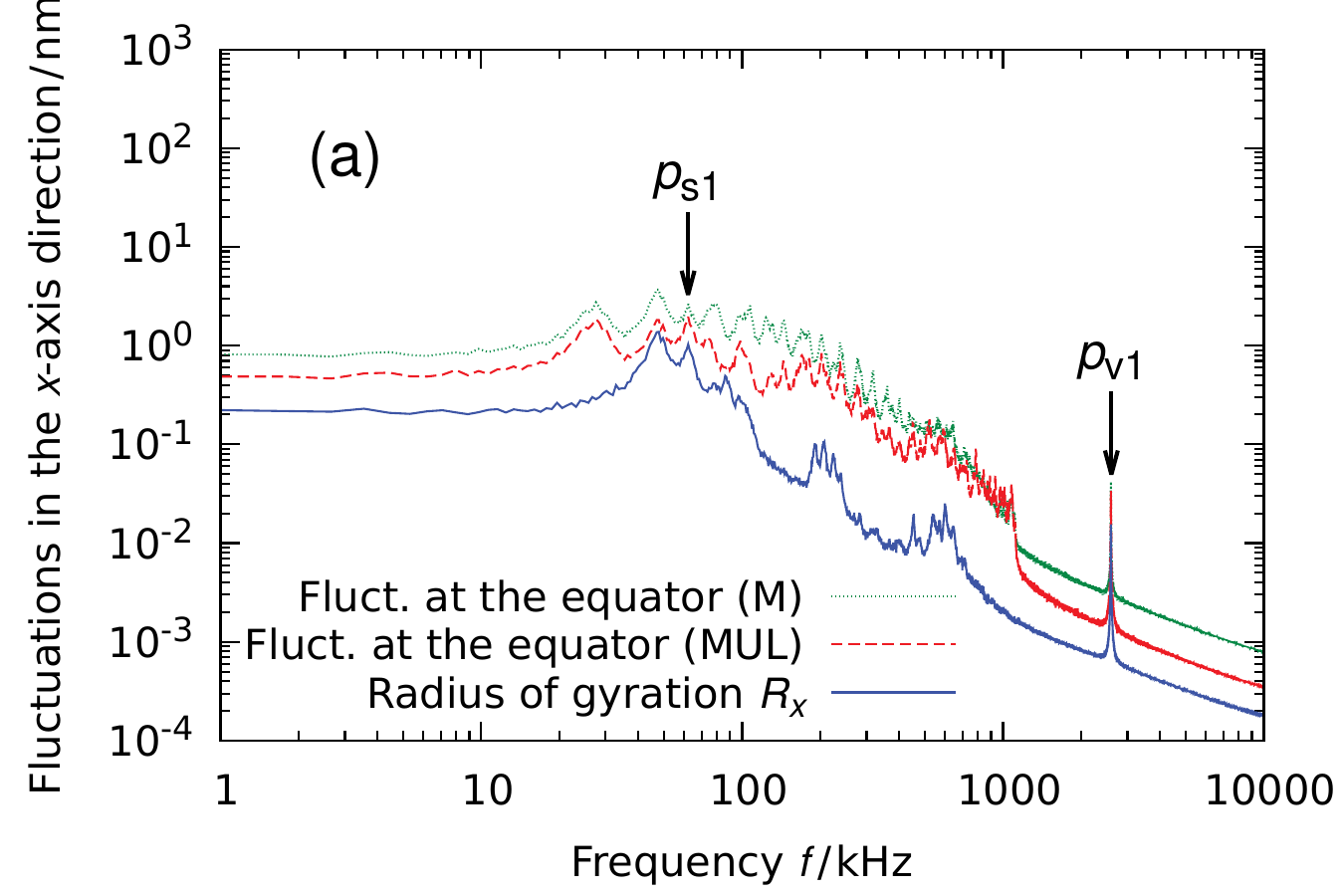}
        \label{subfig:fluct_x}
    \end{subfigure}
    \hfill
    \begin{subfigure}[b]{.49\textwidth}
        \centering
        \includegraphics[width=\linewidth]{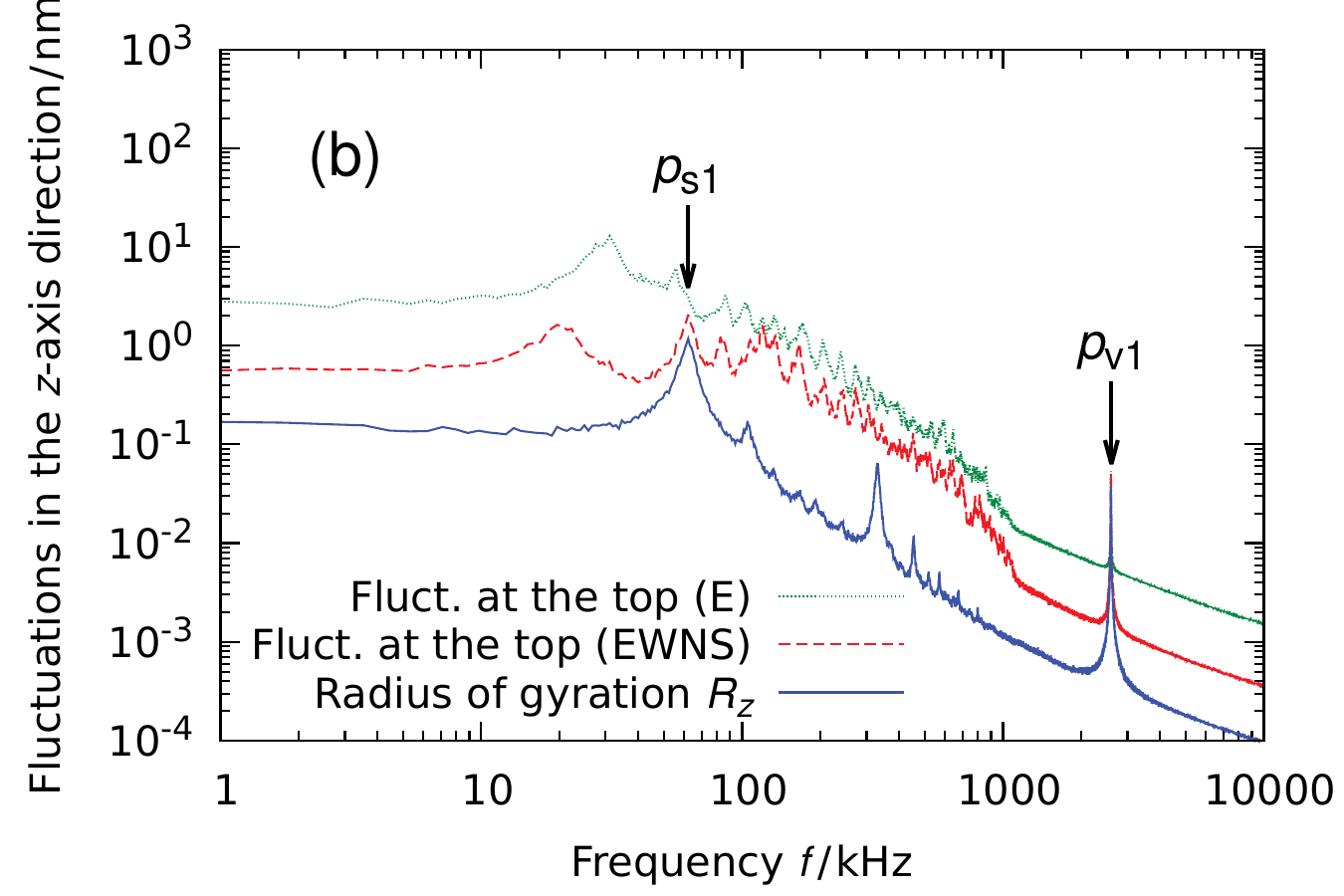}
        \label{subfig:fluct_z}
    \end{subfigure}
    \caption{(Color online) Fourier spectra of the point fluctuations and the radii of gyration $R_x$ and $R_z$ for $\lambda=0.2$. (a) Spectra of fluctuations in the $x$-axis direction. $R_x$ is compared with fluctuations at the middle (M), upper (U), and lower (L) points along the equator. (b) Spectra of fluctuations in the $z$-axis direction. $R_z$ is compared with fluctuations at the east (E), west (W), north (N), and south (S) points at the top. The notations (MUL) and (EWNS) denote the average fluctuations of their respective points.}
    \label{fig:fluct_x_z}
\end{figure}

The comparison between the spectra of point fluctuations and the radii of gyration $R_x$ and $R_z$ are shown in Fig.\ \ref{fig:fluct_x_z} for the damping constant $\lambda=0.2$. The spectra of the fluctuations of particles at the RBC equator are shown in Fig.\ \ref{fig:fluct_x_z}~(a). The notation (MUL) denotes the average of the fluctuations for the three points.
Figure~\ref{fig:fluct_x_z}~(b) shows the spectra of the fluctuations of particles at the top. The notation (EWNS) similarly denotes the average of the fluctuations. In Fig.\ \ref{fig:fluct_x_z}~(a), the peaks $p_\mathrm{s1}$ and $p_\mathrm{v1}$ are observable for all three spectra. However, in Fig.\ \ref{fig:fluct_x_z}~(b), the peak $p_\mathrm{s1}$ is not observed for the fluctuations at the top (E) unlike the fluctuations at the top (EWNS) and $R_z$. This suggests that averaging the membrane fluctuations either over several points or the entire membrane improves the observability of the peaks for fluctuations in the $z$-axis direction.

Although the spectra of the averaged point fluctuations are comparable to those of the radii of gyration, the latter exhibits more distinct peaks in certain frequencies. It should be possible in principle to average the membrane fluctuations over an entire surface (e.g.\ the top surface as seen in Fig.\ \ref{fig:single_points}) similar to the radius of gyration using image analysis techniques. Image analysis has already been employed to measure RBC membrane fluctuations at the rim\cite{yoon2009}. The extension of such methods to the measurement over an entire surface should be considered for future experiments.

\subsubsection{Range of measurable frequencies}

The maximum frequency for the experimentally measured Fourier spectra of RBC membrane fluctuations is approximately $\nunit{100}{Hz}$\cite{humpert2003,gogler2007}. However, the experimental setup proposed by G\"{o}gler \textit{et al.}\ allows us to measure membrane fluctuations up to $\nunit{200}{kHz}$\cite{gogler2007}. Additionally, this maximum frequency is limited only by the bandwidth of the bus connecting the experimental device and the computer. Measurement at higher frequencies around $\nunit{1}{MHz}$ would be feasible for higher-performance buses. In this case, all the peaks found in the present manuscript would be within the range of observable frequencies. Even for the current frequency range capped at $\nunit{200}{kHz}$, the peak $p_\mathrm{s1}$ at $\nunit{62}{kHz}$ is observable. As discussed in the next section, $p_\mathrm{s1}$ is the most important signal. Therefore, measuring $p_\mathrm{s1}$ alone would be meaningful.

\section{Summary and Discussion} \label{sec:summary}

In this study, we developed a method of precisely determining the parameter values used in a DPD model of the RBC by focusing on the fluctuations of the RBC membrane. We measured the Fourier spectra of the FENE potential $V_\mathrm{FENE}$, the area-conserving potential $V_\mathrm{area}$, the volume-conserving potential $V_\mathrm{volume}$, and the bending potential $V_\mathrm{bending}$. Several distinct peaks were observed across multiple spectra. The peaks $p_\mathrm{s1}\ (\nunit{62}{kHz})$ and $p_\mathrm{s2}\ (\nunit{200}{kHz})$ were determined to be from $V_\mathrm{FENE}$, whereas the peaks $p_\mathrm{v1}\ (\nunit{2.6}{MHz})$ and $p_\mathrm{v2}\ (\nunit{5.2}{MHz})$ were found to originate from $V_\mathrm{volume}$.

We further measured the Fourier spectra of the radius of gyration and compared them to the spectra obtained from point fluctuations. The spectra of both measurements exhibited the characteristic peaks $p_\mathrm{s1}$ and $p_\mathrm{v1}$, which reinforces the experimental feasibility of our simulations. The same comparison also suggests that the peaks are better detected for the average of membrane fluctuations at different points than the fluctuations at a single point. We also observed that the peak $p_\mathrm{s1}$ exhibited a singular peak when measured in the $z$-axis direction, making it more detectable than in the $x$-axis direction. We attribute this discrepancy to the anisotropic nature of the RBC membrane, although further studies are necessary to rule out numerical artifacts.

% In interferometric optical tweezer experiments, the membrane displacement is often measured in the $x\text{-}$ and $y\text{-}$axis directions\cite{turlier2016,gogler2007,betz2009}. However, the differences in the spectra of $R_x$ and $R_z$ imply that the frequency of $p_\mathrm{s1}$ is more precisely measured in the $z\text{-}$axis direction.

Theoretically, these results enable us to determine the values of the parameters $k_\mathrm{s}$ and $k_\mathrm{v}$ corresponding to their respective peaks $p_\mathrm{s1}$ and $p_\mathrm{v1}$. In actuality, however, the value of $k_\mathrm{v}$ is limited by the time-step size of the simulation, and that of a physical RBC is much larger than is numerically feasible owing to its highly incompressible nature. On the other hand, the parameter $k_\mathrm{s}$ is determined from the shear modulus of the membrane alone. Therefore, we can determine the value of $k_\mathrm{s}$ by measuring the frequency of $p_\mathrm{s1}$ experimentally.

We believe that the coarse-grained nature of the RBC model used in our simulations exhibits universality to some extent. The potentials $V_\mathrm{area},\ V_\mathrm{volume},\ \text{and }V_\mathrm{bending}$ used in the model reflect the lower-order contributions of the incompressibility and bending energy of the membrane. This means that the equations of these potentials leave little room for alternative formulations. This is not the case for the spring potential $V_\mathrm{spring}$, for which there are multiple appropriate equations. However, regardless of the model equation used, $V_\mathrm{spring}$ can be fundamentally expressed as a quadratic mass-spring-damper (MSD) system around small deviations from equilibrium. In this case, the peak frequencies of the Fourier spectra are determined by the oscillation frequency of the MSD system, irrespective of the details of the spring potential. Note that the inaccuracy of the coarse-grained approximation of the membrane potentials becomes more prominent when the membrane is far from equilibrium such as under large deformations or strong flows\cite{noguchi2009,mcwhirter2009}. Therefore, the scope of the proposed methods is limited to measuring the membrane fluctuations close to equilibrium.

In this study, we considered the lowest oscillation mode of membrane fluctuations. Concurrent measurement in multiple directions will enable us to associate the model parameters with oscillation modes of higher degrees. On the other hand, the membrane fluctuations of an RBC are known to violate the fluctuation--dissipation relation, which suggests the presence of non-equilibrium processes\cite{turlier2016,gnesotto2018}. The methods proposed in the present paper can be applied to investigating the non-equilibrium behavior of RBCs. Measuring the Fourier spectra of the membrane energies and fluctuations will enable us to quantify the dissipation of energy induced by non-equilibrium contributions.

\begin{acknowledgments}
    The authors would like to thank H. Noguchi and H. Nakano for fruitful discussions. This research was supported by JSPS KAKENHI, Grant No.~JP21K11923. The computation was partly carried out using the facilities of the Supercomputer Center, Institute for Solid State Physics (ISSP), University of Tokyo.
\end{acknowledgments}

\appendix

\section{Dependence of peak frequencies on coarse graining} \label{sec:coarse_graining}

The outermost layer of a physical RBC is a lipid bilayer with the cytoskeleton attached underneath\cite{nir2007}. The lipid bilayer is continuous in the length scale of an RBC, whereas the cytoskeleton is estimated to contain approximately $27000\,\text{--}\, 45000$ actin nodes\cite{fedosov2010}. This is on the basis of a node density of $200\,\text{--}\, 330/\mathrm{\mu m^2}$, with $\nunit{135}{\mu m^2}$ adopted as the average surface area\cite{takeuchi1998,swihart2001}. On the other hand, the actin nodes are coarse-grained using 492 particles in this study, which is considerably less than the actual values. Therefore, we must investigate how the coarse graining affects the peak frequencies of the Fourier spectra. We performed a simulation varying the number of particles used to discretize the RBC. We measured the frequencies of the aforementioned peaks $p_\mathrm{s1}$ of $V_\mathrm{FENE}$ and $p_\mathrm{v1}$ of $V_\mathrm{volume}$ with the number of particles $N=162,\,252,\,362,\,492,\,\text{and}\,1002$. The same values of the model parameters $k_\mathrm{s},\ k_\mathrm{a},\ k_\mathrm{v}$, and $k_\mathrm{b}$ were used in each case, independent of $N$. On the other hand, the particle mass was adjusted to keep the membrane mass constant.

\begin{figure}[htbp]
    \begin{subfigure}[htbp]{.49\textwidth}
        \centering
        \includegraphics[width=\linewidth]{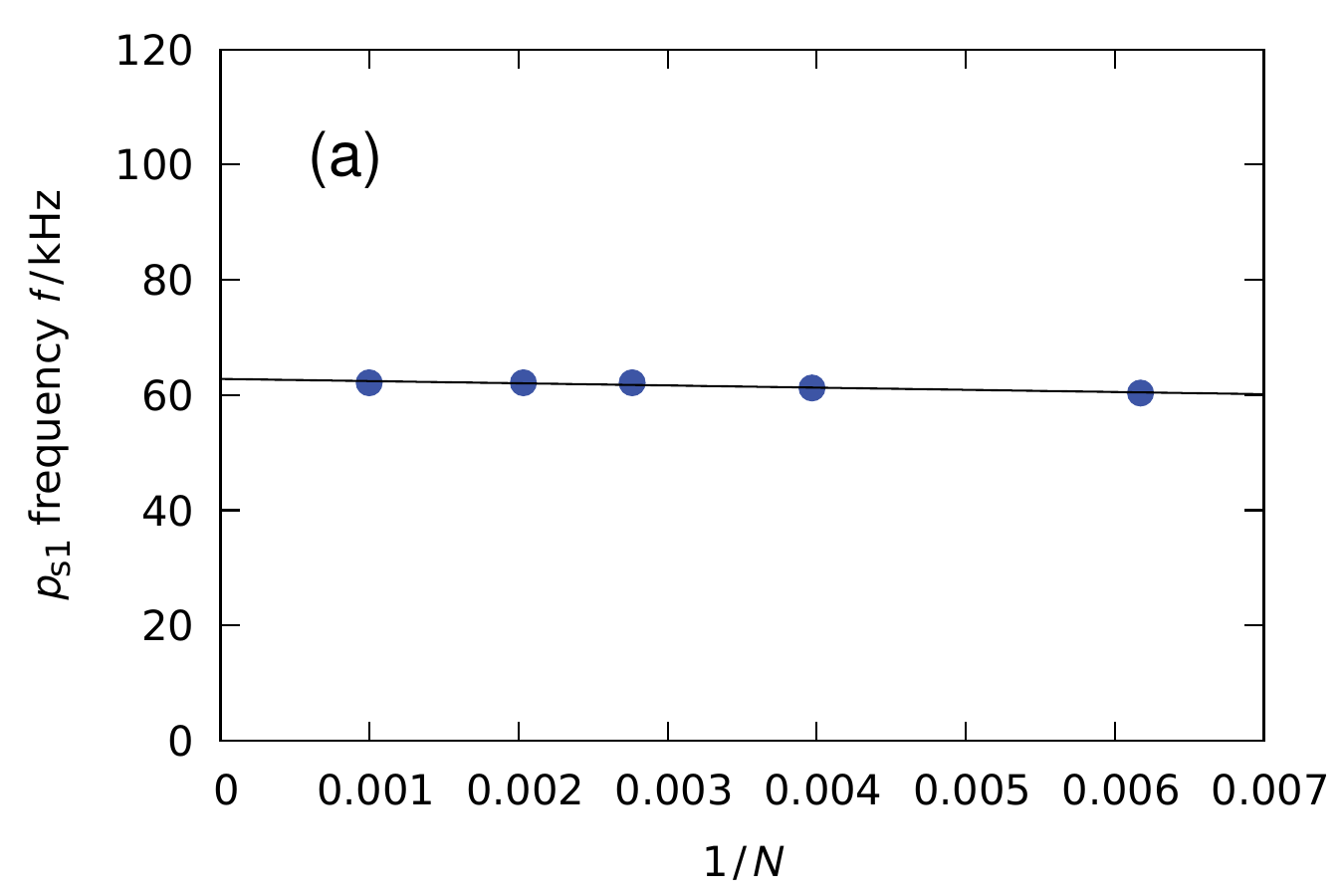}
        \label{subfig:ps1}
    \end{subfigure}
    \hfill
    \begin{subfigure}[htbp]{.49\textwidth}
        \centering
        \includegraphics[width=\linewidth]{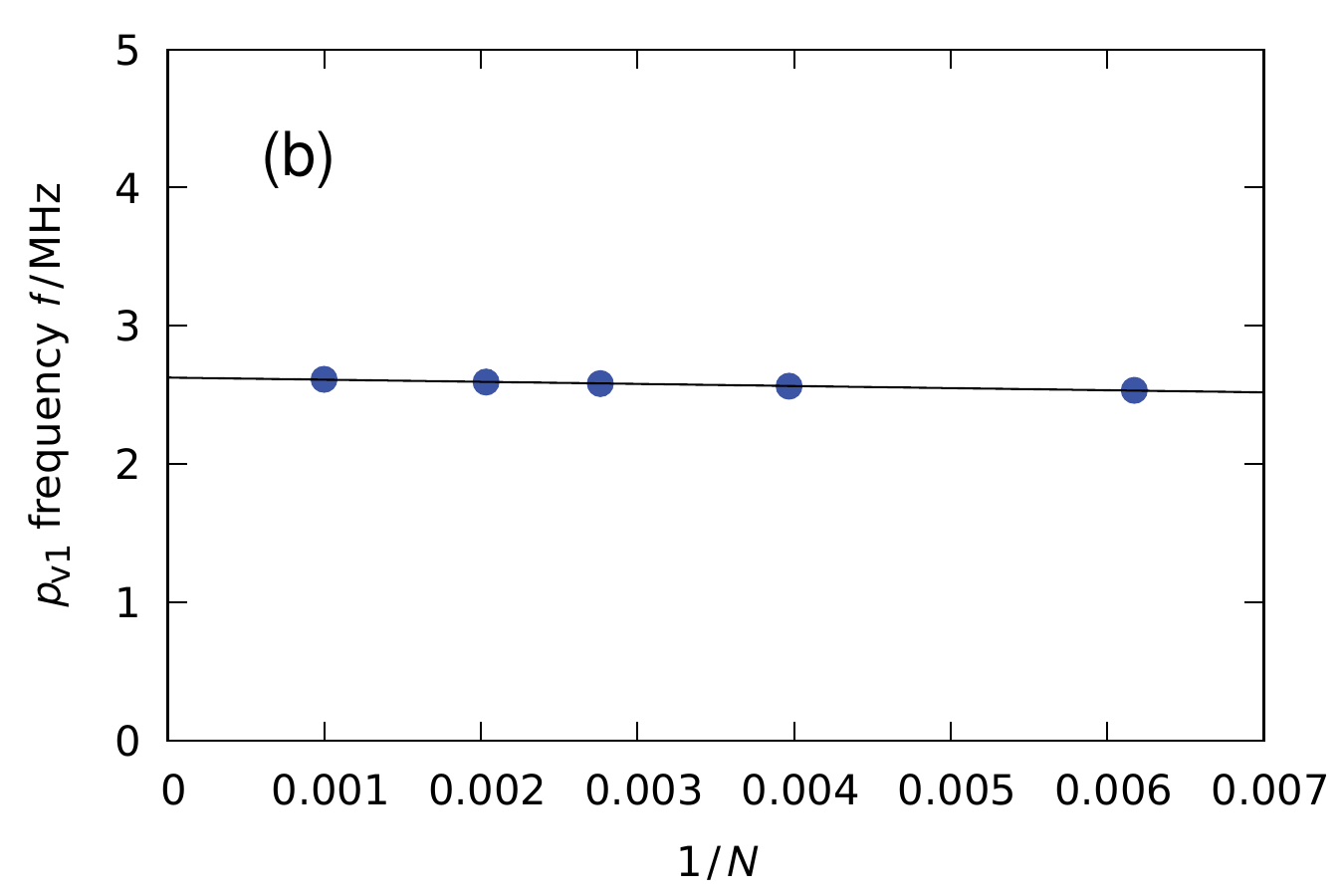}
        \label{subfig:pv1}
    \end{subfigure}
    \caption{Peak frequencies plotted against $1/N$, where $N$ is the number of particles: (a) $p_\mathrm{s1}$ and (b) $p_\mathrm{v1}$. Plot points are fit to $f(N)=f(\infty)+a/N$, where $f(N)$ denotes the peak frequency.}
    \label{fig:ps1_pv1}
\end{figure}

As shown in Fig.~\ref{fig:ps1_pv1}, the peak frequencies of $p_\mathrm{s1}$ and $p_\mathrm{v1}$ exhibit virtually no dependence on the number of particles. The plot points were fit to $f(N)=f(\infty)+a/N$, where $f(N)$ denotes the peak frequency. By comparing the peak frequencies at $N\to\infty$ (the continuum limit) and $N=492$ (the number of particles used in the simulations), we found that the difference between $f(\infty)$ and $f(492)$ was within $1\%$ in both Figs.~\ref{fig:ps1_pv1} (a) and (b). Therefore, the dependence of peak frequencies on coarse graining is negligible for the peaks of both $V_\mathrm{FENE}$ and $V_\mathrm{volume}$.

\section{Equations of membrane forces} \label{sec:forces}

The nodal forces corresponding to the membrane potentials $V_\mathrm{spring},\ V_\mathrm{area},\ V_\mathrm{volume}$, and $V_\mathrm{bending}$ will be explained below. Definitions and notations are mostly identical to those in the previous work by Fedosov\cite{fedosov2010phd}.

\subsection{Spring forces from $V_\mathrm{spring}$} \label{subsec:two_body_forces}

The force $\boldsymbol{F}_{ij}^\mathrm{spring}$ corresponding to Eq.~\eqref{eq:Vspring} is a force that the $j$th particle exerts on the neighboring $i$th particle along the side of a triangular lattice:
\begin{equation}
    \label{eq:Fspring}
    \boldsymbol{F}_{ij}^\mathrm{spring}=\left(-\dfrac{k_\mathrm{s} l_{ij}}{1-x^2}+\dfrac{k_\mathrm{p}}{l_{ij}^2}\right)\hat{\boldsymbol{l}}_{ij}.
\end{equation}
Here, $l_{ij}$ is the length of the spring and $x=l_{ij}/l_{ij}^\mathrm{m}$, where $l_{ij}^\mathrm{m}$ is the maximum length of the spring. Additionally, $\hat{\boldsymbol{l}}_{ij}=\boldsymbol{l}_{ij}/l_{ij}$ is a unit vector between the two ends of the spring.

\subsection{Area- and volume-conserving forces from $V_\mathrm{area}$ and $V_\mathrm{volume}$} \label{subsec:three_body_forces}

\begin{figure}[htbp]
    \includegraphics[width=.5\textwidth]{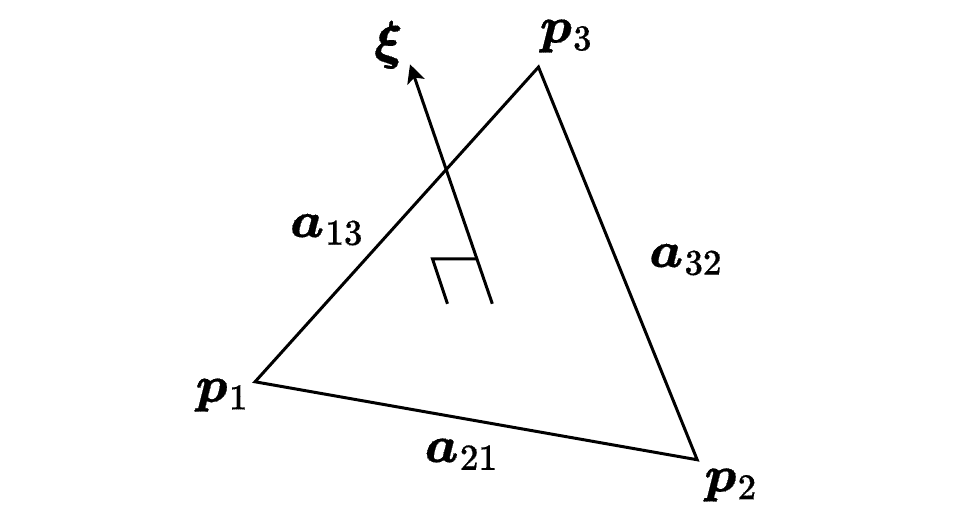}
    \caption{Illustration of a triangular lattice comprising the RBC membrane. $\boldsymbol{p}_i$ represents a vertex and $\boldsymbol{a}_{ij}$ denotes a side, where $i,j=1,2,3$. $\boldsymbol{\xi}$ is a normal vector always taken to point outward from the membrane.}
    \label{fig:three-body}
\end{figure}

Suppose we take a single triangular lattice comprising the RBC membrane, as shown in Fig.~\ref{fig:three-body}. The vector extending from the $j$th particle to the $i$th particle is denoted by $\boldsymbol{a}_{ij}=\boldsymbol{p}_i-\boldsymbol{p}_j$, where $i,j=1,2,3$. Furthermore, the normal vector $\boldsymbol{\xi}=\boldsymbol{a}_{21}\times \boldsymbol{a}_{31}$ is taken such that it always points outward away from the membrane. The area $A_t$ and the volume $V_t$ occupied by the current lattice at time $t$ are expressed as $A_t=\norm{\boldsymbol{\xi}}/2$ and $V_t=\boldsymbol{\xi}\cdot\boldsymbol{r}_\mathrm{c}/6$. Here, $\boldsymbol{r}_\mathrm{c}=(\boldsymbol{p}_1+\boldsymbol{p}_2+\boldsymbol{p}_3)/3$ is the center of mass of the lattice relative to that of the entire membrane. Given these definitions, consider the following coefficients $\beta_\mathrm{a}$ and $\beta_\mathrm{v}$:
\begin{equation}
    \beta_\mathrm{a}=-k_\mathrm{a}\frac{A_t^\mathrm{tot}-A_0^\mathrm{tot}}{A_0^\mathrm{tot}},\quad
    \beta_\mathrm{v}=-k_\mathrm{v}\frac{V_t^\mathrm{tot}-V_0^\mathrm{tot}}{V_0^\mathrm{tot}},
\end{equation}
where $A_t^\mathrm{tot}$ and $A_0^\mathrm{tot}$ are the current and initial total membrane areas, respectively, with an analogous notation for the volumes $V_t^\mathrm{tot}$ and $V_0^\mathrm{tot}$. Using $\beta_\mathrm{a}$ and $\beta_\mathrm{v}$, we can write the forces $\boldsymbol{F}_i^\mathrm{area}$ and $\boldsymbol{F}_i^\mathrm{volume}$ acting on the $i$th $(i=1,2,3)$ particle of the triangular lattice as
\begin{align}
    \boldsymbol{F}_1^\mathrm{area}=\frac{\beta_\mathrm{a}}{4A_t}\left(\boldsymbol{\xi}\times\boldsymbol{a}_{32}\right), & \quad\boldsymbol{F}_1^\mathrm{volume}=\frac{\beta_\mathrm{v}}{6}\left(\frac{\boldsymbol{\xi}}{3}+\boldsymbol{r}_\mathrm{c}\times\boldsymbol{a}_{32}\right), \\
    \boldsymbol{F}_2^\mathrm{area}=\frac{\beta_\mathrm{a}}{4A_t}\left(\boldsymbol{\xi}\times\boldsymbol{a}_{13}\right), & \quad\boldsymbol{F}_2^\mathrm{volume}=\frac{\beta_\mathrm{v}}{6}\left(\frac{\boldsymbol{\xi}}{3}+\boldsymbol{r}_\mathrm{c}\times\boldsymbol{a}_{13}\right), \\
    \boldsymbol{F}_3^\mathrm{area}=\frac{\beta_\mathrm{a}}{4A_t}\left(\boldsymbol{\xi}\times\boldsymbol{a}_{21}\right), & \quad\boldsymbol{F}_3^\mathrm{volume}=\frac{\beta_\mathrm{v}}{6}\left(\frac{\boldsymbol{\xi}}{3}+\boldsymbol{r}_\mathrm{c}\times\boldsymbol{a}_{21}\right),
\end{align}
where $\boldsymbol{F}_i^\mathrm{area}$ and $\boldsymbol{F}_i^\mathrm{volume}$ correspond to Eqs. \eqref{eq:Varea} and \eqref{eq:Vvolume}, respectively.

\subsection{Bending forces from $V_\mathrm{bending}$} \label{subsec:four_body_forces}

\begin{figure}[htbp]
    \includegraphics[width=.6\textwidth]{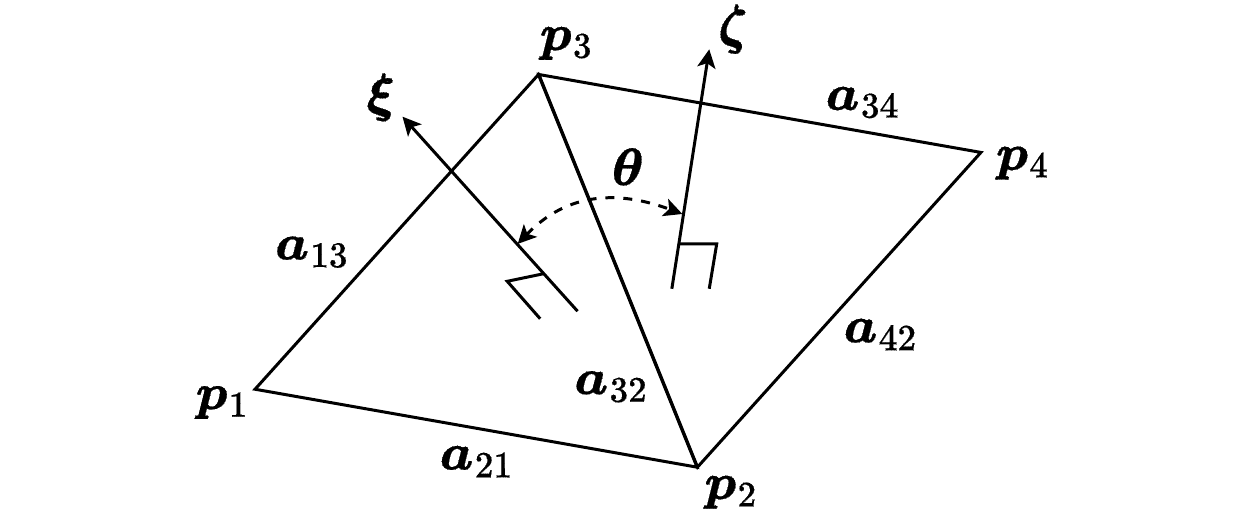}
    \caption{Illustration of two adjacent triangular lattices taken from the RBC membrane. $\boldsymbol{p}_i$ represents a vertex and $\boldsymbol{a}_{ij}$ denotes a side, where $i,j=1,2,3,4$. $\boldsymbol{\xi}$ and $\boldsymbol{\zeta}$ are normal vectors, whereas $\theta$ is the dihedral angle equal to the angle between $\boldsymbol{\xi}$ and $\boldsymbol{\zeta}$.}
    \label{fig:four-body}
\end{figure}

Shown in Fig.~\ref{fig:four-body} are two adjacent triangular lattices, where the normal vectors $\boldsymbol{\xi}=\boldsymbol{a}_{21}\times\boldsymbol{a}_{31}$ and $\boldsymbol{\zeta}=\boldsymbol{a}_{34}\times\boldsymbol{a}_{24}$. If $(\boldsymbol{\xi}-\boldsymbol{\zeta})\cdot(\boldsymbol{r}_\mathrm{c}^{\boldsymbol{\xi}}-\boldsymbol{r}_\mathrm{c}^{\boldsymbol{\zeta}})<0$, where $\boldsymbol{r}_\mathrm{c}^{\boldsymbol{\xi}}$ and $\boldsymbol{r}_\mathrm{c}^{\boldsymbol{\zeta}}$ are the centers of mass of the respective lattices, then the labels $\boldsymbol{p}_2$ and $\boldsymbol{p}_3$ are swapped, after which the sides and normal vectors are recalculated. The dihedral angle $\theta$ is written as
\begin{equation}
    \theta=\cos^{-1}\left(\frac{\boldsymbol{\xi}}{\norm{\boldsymbol{\xi}}}\cdot\frac{\boldsymbol{\zeta}}{\norm{\boldsymbol{\zeta}}}\right),
\end{equation}
which is equal to the angle between $\boldsymbol{\xi}$ and $\boldsymbol{\zeta}$. We then define three coefficients,
\begin{equation}
    b_{11}=-\frac{\beta_\mathrm{b}\cos\theta}{\norm{\boldsymbol{\xi}}^2},\quad b_{12}=\frac{\beta_\mathrm{b}}{\norm{\boldsymbol{\xi}}\norm{\boldsymbol{\zeta}}},\quad b_{22}=-\frac{\beta_\mathrm{b}\cos\theta}{\norm{\boldsymbol{\zeta}}^2},
\end{equation}
where $\beta_\mathrm{b}=k_\mathrm{b}\sin(\theta-\theta_0)/\sqrt{1-\cos^2\theta}$, with $\theta_0$ being the spontaneous angle. These definitions provide a force $\boldsymbol{F}_i^\mathrm{bending}$ corresponding to Eq.~\eqref{eq:Vbending}, exerted on the $i$th particle $(i=1,2,3,4)$ of the two adjacent triangular lattices:
\begin{align}
    \boldsymbol{F}_1^\mathrm{bending} & =b_{11}\left(\boldsymbol{\xi}\times\boldsymbol{a}_{32}\right)+b_{12}\left(\boldsymbol{\zeta}\times\boldsymbol{a}_{32}\right),                                                                                                          \\
    \boldsymbol{F}_2^\mathrm{bending} & =b_{11}\left(\boldsymbol{\xi}\times\boldsymbol{a}_{13}\right)+b_{12}\left(\boldsymbol{\xi}\times\boldsymbol{a}_{34}+\boldsymbol{\zeta}\times\boldsymbol{a}_{13}\right)+b_{22}\left(\boldsymbol{\zeta}\times\boldsymbol{a}_{34}\right), \\
    \boldsymbol{F}_3^\mathrm{bending} & =b_{11}\left(\boldsymbol{\xi}\times\boldsymbol{a}_{21}\right)+b_{12}\left(\boldsymbol{\xi}\times\boldsymbol{a}_{42}+\boldsymbol{\zeta}\times\boldsymbol{a}_{21}\right)+b_{22}\left(\boldsymbol{\zeta}\times\boldsymbol{a}_{42}\right), \\
    \boldsymbol{F}_4^\mathrm{bending} & =b_{12}\left(\boldsymbol{\xi}\times\boldsymbol{a}_{23}\right)+b_{22}\left(\boldsymbol{\zeta}\times\boldsymbol{a}_{23}\right).
\end{align}

\section{Spectra in the NVT ensemble} \label{sec:dpd_on_spectra}

\begin{figure}[htbp]
    \begin{subfigure}[htbp]{.49\textwidth}
        \centering
        \includegraphics[width=\linewidth]{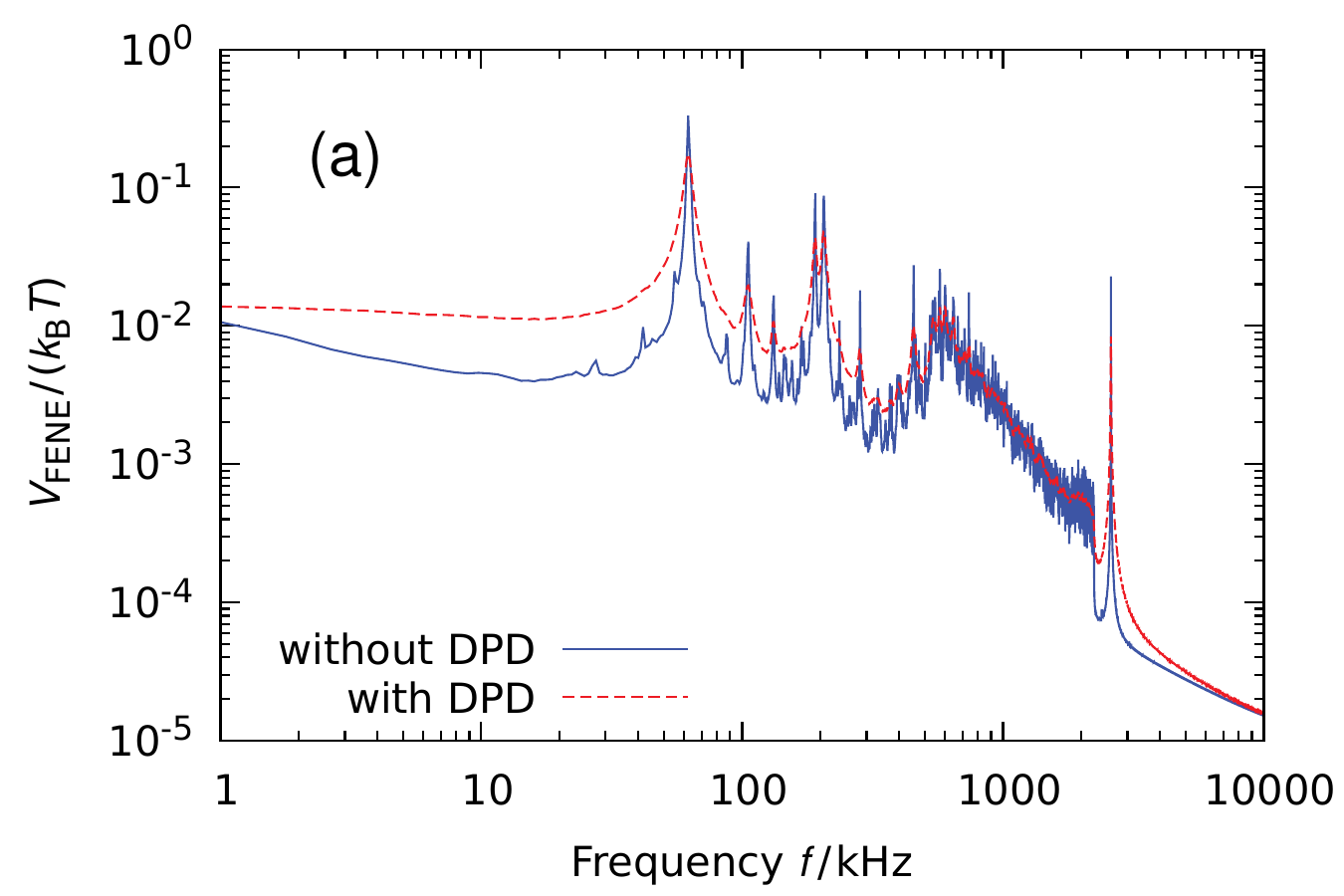}
        \label{subfig:dpd_spring}
    \end{subfigure}
    \hfill
    \begin{subfigure}[htbp]{.49\textwidth}
        \centering
        \includegraphics[width=\linewidth]{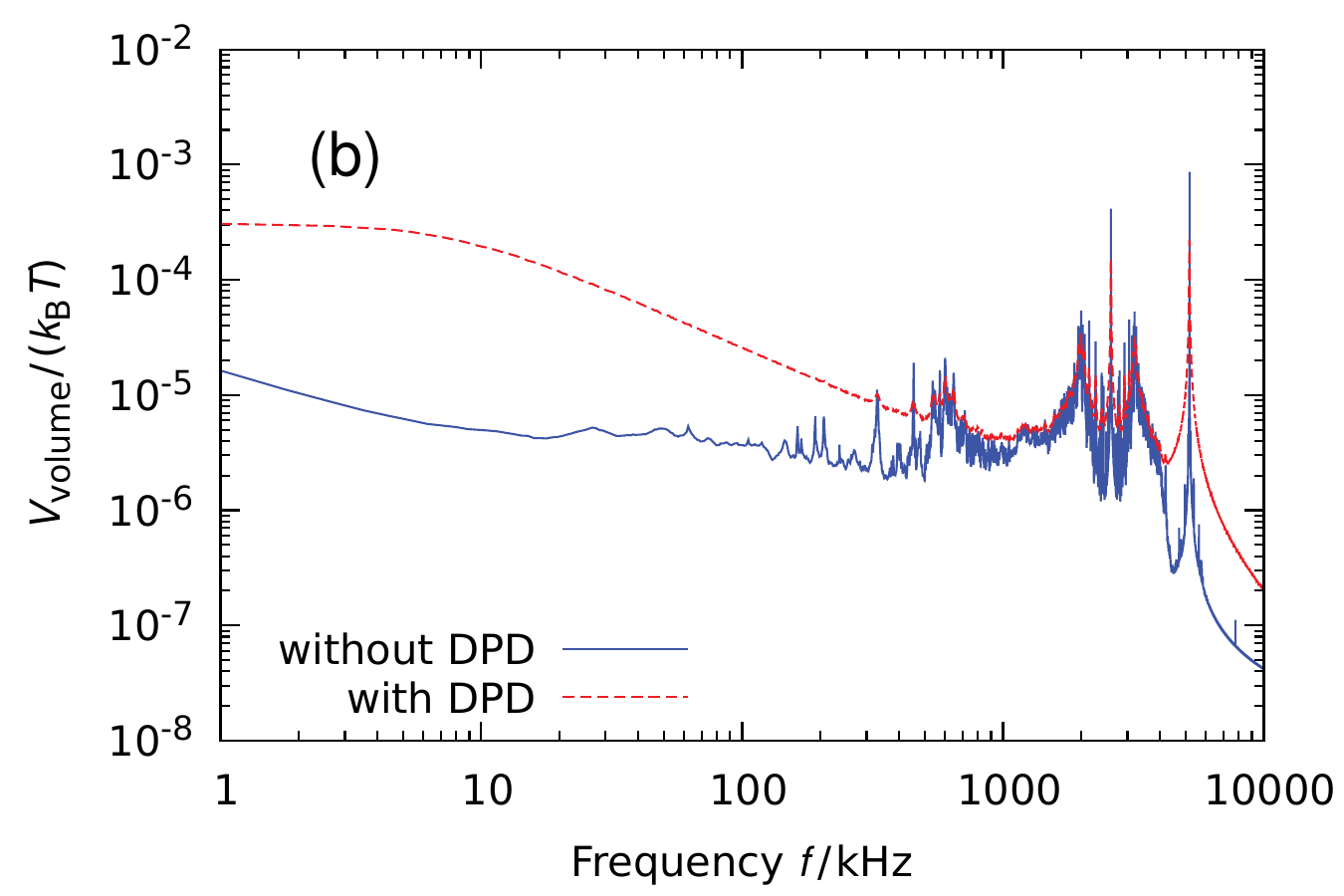}
        \label{subfig:dpd_volume}
    \end{subfigure}
    \caption{(Color online) Fourier spectra of fluctuations of membrane potentials: (a) $V_\mathrm{FENE}$ and (b) $V_\mathrm{volume}$. The spectra in dashed lines are those of an NVT simulation, and they are superimposed on the solid spectra of the NVE simulation discussed in Sec.~\ref{sec:results}.}
    \label{fig:dpd_w_wo}
\end{figure}

The Fourier spectra listed in Sec.~\ref{sec:results} of this paper were obtained from an NVE simulation where the DPD thermostat was turned off after the membrane reached the equilibrium. As a comparison, Fig.~\ref{fig:dpd_w_wo} shows in dashed lines the spectra of $V_\mathrm{FENE}$ and $V_\mathrm{volume}$ from an NVT simulation where the thermostat was applied throughout. The corresponding spectra from the NVE simulation are shown in solid lines. The two spectra differ in that the NVT spectra have a less pronounced profile, have wider peaks, and show a bias at lower frequencies. Crucially, however, all peak frequencies remain unchanged between NVT and NVE. Therefore, we opted to measure the Fourier spectra in NVE to study the spectra and their peaks in more detail.

\bibliography{reference}

\end{document}